\documentclass[superscriptaddress,twocolumn,floatfix,amsmath,amssymb,pra]{revtex4}

\usepackage{amsmath}  
\usepackage{graphicx} 
\usepackage{verbatim} 
\usepackage{color}   
\usepackage{subfigure}
\usepackage{amssymb}
\usepackage{amsmath}
\usepackage{mathrsfs}
\usepackage{amsfonts}
\usepackage{braket}
\usepackage{appendix}
\usepackage{epstopdf}

\DeclareMathAlphabet{\mathpzc}{OT1}{pzc}{m}{it}

\begin{document}

\title{Quantum Defect Theory for High Partial Wave Cold Collisions}

\author{Brandon P. Ruzic}
\affiliation{JILA, University of Colorado and National Institute of Standards and Technology, Boulder, Colorado 80309-0440, USA}
\author{Chris H. Greene}
\affiliation{JILA, University of Colorado and National Institute of Standards and Technology, Boulder, Colorado 80309-0440, USA}
\affiliation{Department of Physics, Purdue University, West Lafayette, IN 47907}
\author{John L. Bohn}
\affiliation{JILA, University of Colorado and National Institute of Standards and Technology, Boulder, Colorado 80309-0440, USA}

\begin{abstract} We extend Multichannel Quantum Defect Theory (MQDT) to ultracold collisions involving high partial wave quantum numbers $L$. This requires a careful standardization of the MQDT reference wave functions at long range to ensure their linear independence. To illustrate the simplicity and accuracy of the resulting theory, we perform a comprehensive calculation of $L\le 2$ Fano-Feshbach resonances in the range $0-1000$ G for the scattering of $^{40}$K + $^{87}$Rb in their lowest hyperfine states. 
\end{abstract}


\maketitle

\section{Introduction}

The constituents of cold gaseous matter continue to
grow in variety and complexity far beyond their origins in alkali-metal atoms to
encompass open shell atoms, molecules, free radicals, and ions
\cite{1367-2630-11-5-055009}.  A detailed understanding of their collision
processes is crucial in determining the properties and
prospects for control of these gases.

However, along with the growing complexity of the atoms and molecules
involved, the difficulty of accurate scattering calculations grows as well.
High-spin and open-shell atoms contain a multiplicity of internal states, and
molecules incorporate rotational and vibrational degrees of freedom.
At ultralow temperatures all these degrees of freedom must be accounted for because they all describe energies that are typically large compared to translational kinetic energies in the gas.  Moreover, anisotropic species are likely to involve angular momentum partial waves far larger than
the single value $L=0$ that often dominates alkali atom cold collisions.
These considerations can lead to enormous complexity, even in seemingly
straightforward problems. Consider for example the scattering
of Li atoms with ground state Li$_2$ molecules, which necessitates
the simultaneous solution of thousands of coupled Schr\"{o}dinger equations \cite{Quemener2007}.

It is therefore worthwhile to explore alternative methods of scattering
theory that are less computationally intensive, yet still accurate.
A particularly appealing
candidate for this purpose is the Multichannel Quantum Defect Theory (MQDT),
which was originally developed to describe complex spectra
of atoms \cite{Schroodinger1921} \cite{0034-4885-46-2-002} \cite{Fano1986}.  It has more recently seen fruitful application
to atomic collisions \cite{mies:2514} and, in particular, cold collisions \cite{Fourre1994} \cite{PhysRevA.70.012710} \cite{PhysRevA.84.042703}.

In recent years various approaches have been applied to ultracold
collisions with an aim of describing and predicting scattering
observables and, in particular, the locations and widths of
Fano-Feshbach resonances.  For example, recent developments in MQDT
and related ideas have stressed the simplicity and analytic behavior of
the theory \cite{PhysRevLett.81.3355} \cite{PhysRevA.72.042719} \cite{PhysRevA.74.052715} \cite{gao:012702} \cite{hanna:040701} \cite{julienne09} \cite{PhysRevLett.104.113202} \cite{PhysRevA.81.022702} \cite{PhysRevA.82.020703}. Also, the analysis of long-range s-wave solutions developed in \cite{PhysRevA.84.032701} adopts semiclassical ideas and creates a good approximation to the long-range field solutions that resembles quantum defect ideas to a degree. To a good approximation, three parameters can describe cold collisions of alkali-metal atoms. There are two scattering lengths that describe the physics at small interparticle separation $R$ and one dispersion coefficient $C_6$ that describes the physics at large $R$ \cite{hanna:040701}. In parallel developments, the asymptotic bound state method stresses direct numerical diagonalization in a basis of singlet and triplet states that are coupled by hyperfine and magnetic interactions \cite{wille:053201} \cite{PhysRevLett.104.053202} \cite{PhysRevA.82.042712}. This has generated extremely rapid and accurate numerical calculations in these cases.

The present paper describes an attempt to retain the analytic structure of MQDT and simultaneously perform numerically exact scattering calculations.  On the one hand, MQDT allows the ``hard'' short-range physics to be described via parameters
that are nearly energy- and field-independent. In addition, analytic expressions are derived for some of the long-range parameters in the threshold limit.  On the other hand, our treatment is numerically exact, even in the case of realistic long-range
potentials that are not purely of $-C_6/R^6$ character and that possess
high centrifugal angular momentum. As a numerical method, it is also fast because it eschews repeated numerical calculation of the short-range wave functions.

MQDT takes into account the natural separation of length and energy scales
inherent in collision problems.  Namely, the sensitive dependence of collision
observables on energy and electric or magnetic fields arises from
interactions between the scattering partners at large $R$. These dependencies
 are described ultimately through
a carefully chosen set of reference wave functions in this region.  This part
of the calculation is relatively fast, as the reference wave functions are defined
separately for each set of quantum numbers identifying a scattering channel.
Vice versa, the ``hard'' part of the calculation lies at smaller $R$,
where the channels
are strongly coupled together and must all be dealt with at once.
However, in this circumstance, the energy scales driving the physics are
far larger than the $\mu$K - mK
scale of cold collisions.  A properly chosen representation
of the small-$R$ wave function can then be quite weakly energy dependent,
allowing ready interpolation in energy and field that greatly reduces
computational time \cite{PhysRevLett.81.3355}.

However, accuracy of the MQDT method relies on the availability of a pair of
accurate, linearly independent reference wave functions in each channel.
This can become problematic in cases of high partial wave angular momentum and
extremely low collision energy, where there exists a substantial
region of classically forbidden motion underneath a centrifugal
barrier.  Such a barrier
is problematic because two reference wave functions that are perfectly
linearly independent
prior to entering the barrier can often become approximately linearly dependent
(hence useless to the theory) underneath and beyond this barrier.

Our main focus here is to determine reference wave
functions in a way that maximizes their linear independence, even underneath
the centrifugal barrier. The method is then applied to the scattering of potassium and rubidium atoms, for which a reasonable and experimentally constructed Hamiltonian already exists and the long-range interaction includes contributions from $C_8$ and $C_{10}$ in addition to the usual $C_6$. MQDT reproduces the full close-coupling calculation quite accurately with orders of magnitude less computational effort. In particular, this version of MQDT reproduces magnetic field Fano-Feshbach resonances accurately, even those that reside in high partial wave states.

\section{Theory}
\subsection{Scattering in the MQDT Picture}
In two-body atomic or molecular scattering, the wave function can be expanded in a basis of $N$ magnetic-field-dressed hyperfine channels that include the partial wave quantum number $L_i$,
\begin{equation}
 \psi=R^{-1}\sum_{i=1}^N\Phi_i(\Omega)\psi_i(R)\text{ },
\end{equation}
where $\Omega$ represents all angular coordinates and spin degrees of freedom. This wave function satisfies a set of coupled radial Schr\"{o}dinger equations involving the potential matrix $V(R)$,
\begin{equation}
 \label{eq:SE}
 \sum_{j=1}^N\left[\left(-\frac{\mathrm{d}^2}{\mathrm{d} R^2}+\frac{L_j(L_j+1)}{R^2}\right)\delta_{ij}+V_{ij}\right]\psi_j=E_i\psi_i\text{ }.
\end{equation}
$E_i$ is the channel collision energy, i.e., the energy above threshold for each channel $E_i=E-E^{\text{thresh}}_i$, where $E$ is the collision energy and $E^{\text{thresh}}_i$ is the threshold energy of the $i$-th channel.

Here and throughout this paper (unless otherwise specified), all lengths are in units of the natural length scale $\beta$ of the potential $V$, and all energies are in units of the natural energy scale $E_{\beta}=\hbar^2/2\mu\beta^2$, where $\mu$ is the reduced mass. As a consequence, $\mu$ is scaled out of many equations, and $k_i=\sqrt{E_i-V_{ij}}$ has units of $\beta^{-1}$. The form of $V$ at long range suggests the value of $\beta$. For example, if $V$ is well described at long range by the isotropic potential $-C_6/R^6$, the natural unit of length is $\beta=(2\mu C_6/\hbar^2)^{1/4}$. Later sections discuss the collision of $^{40}$K + $^{87}$Rb, for which $\beta=143.9$ a$_0$ and $E_\beta=152.7$ $\mu$K, where a$_0$ is the Bohr radius.

One can calculate scattering observables by solving equation (\ref{eq:SE}) subject to physical boundary conditions. In each channel $\psi$ has the boundary conditions $\psi_i=0$ at $R=0$. The closed-channel $(E_i<0)$ components of $\psi$ vanish at $R=\infty$. Hence, if $N_o$ is the number of open channels $(E_i>0)$, there exist $N_o$ independent $\psi$'s which only have a non-vanishing amplitude in the open channels asymptotically. The open-channel components of these wave functions are energy-normalized and represent the $N_o\times N_o$ asymptotic solution matrix $\Psi$.

Since the potential goes to zero in the limit $R\rightarrow\infty$, the open-channel wave functions become linear combinations of sine and cosine, and $\Psi$ can be written in terms of energy-normalized, free-particle solutions,
\begin{equation}
  \label{eq:psi}
  \Psi_{ij}\xrightarrow{R\rightarrow \infty}\frac{1}{\sqrt{k_i}}(k_iR)\left(j_{L_i}(k_iR)\delta_{ij}-n_{L_i}(k_iR)K_{ij}^{\text{phys}}\right)\text{ }, 
\end{equation}
where $j_L$ and $n_L$ are spherical Bessel functions of the first and second kind, respectively,
\begin{subequations}
\label{eq:bess}
\begin{align}
 j_{L_i}(k_iR)&\xrightarrow{R\rightarrow \infty} \frac{\sin(k_iR-L_i\pi/2)}{k_iR}\text{ },\\
 n_{L_i}(k_iR)&\xrightarrow{R\rightarrow \infty} -\frac{\cos(k_iR-L_i\pi/2)}{k_iR}\text{ }.
\end{align}
\end{subequations}
Here, $\delta_{ij}$ is the Kronecker delta function, and $k_i=\sqrt{E_i-V_{ii}}\xrightarrow{R\rightarrow\infty}\sqrt{E_i}$. Equation (\ref{eq:psi}) defines the physical $K$-matrix which contains all the information necessary to compute scattering observables, including resonance behavior and threshold effects, and is simply related to the scattering matrix,
\begin{equation}
 S^{\text{phys}}=(I+iK^{\text{phys}})(I-iK^{\text{phys}})^{-1}\text{ },
\end{equation}
where $I$ is the identity matrix.
  
For many problems, such as atoms and molecules interacting via van der Waals potentials, the channels are approximately uncoupled beyond a radius $R_\text{m}$. In general, there are $N$ independent solutions to equation (\ref{eq:SE}) that have only the boundary conditions $\psi_i=0$ at $R=0$ and represent the $N\times N$ solution matrix $M$. Hence, matching $M$ to single-channel reference wave functions $\hat{f}$ and $\hat{g}$ at $R=R_\text{m}$ defines a short-range $K$-matrix \cite{RevModPhys.68.1015},
\begin{equation}
  M_{ij}=\hat{f}_i\delta_{ij}-\hat{g}_iK_{ij}^{\text{sr}}\text{ }. 
\end{equation}
Here, $\hat{f}$ and $\hat{g}$ are solutions to the uncoupled radial Schr\"{o}dinger equations in the long-range potential $V^{\text{lr}}$,
\begin{equation}
 \label{eq:ref}
 \left(-\frac{\mathrm{d}^2}{\mathrm{d} R^2}+\frac{L_i(L_i+1)}{R^2}+V_i^{\text{lr}}-E_i\right)
\begin{Bmatrix}
\ \hat{f}_i \\
\ \hat{g}_i \\
\end{Bmatrix} 
 = 0\text{ }.
\end{equation}
The matching is best done when all channels are locally open $\left(E_i>V_{ij}(R_\text{m})\right)$ because $\hat{f}$ and $\hat{g}$ are oscillatory and can, therefore, easily be made linearly independent. To the extent that the channel coupling is negligible beyond $R_\text{m}$, applying boundary conditions at $R=\infty$ in terms of $\hat{f}$ and $\hat{g}$ allows scattering observables to be computed accurately.

\subsection{Reference Wave Functions}
The major benefit of using the reference wave functions $\hat{f}$ and $\hat{g}$ is that they do not need to satisfy physical boundary conditions. Their boundary conditions can instead be smooth, analytic functions of energy. In particular, choosing WKB boundary conditions well within the classically allowed region at $R=R_\text{x}\le R_\text{m}$ accomplishes this goal \cite{mies:2514} \cite{PhysRevA.30.3321},
\begin{subequations}
\label{eq:fhatandghat}
\begin{align}
\label{eq:fhat}
\hat{f}_i(R)&=\frac{1}{\sqrt{k_i(R)}}\sin(\int_{R_\text{x}}^{R} k_i(R') \mathrm{d}R' + \phi_i) \quad\text{at }R=R_\text{x}\text{ },\\
\label{eq:ghat}
\hat{g}_i(R)&=-\frac{1}{\sqrt{k_i(R)}}\cos(\int_{R_\text{x}}^{R} k_i(R') \mathrm{d}R' + \phi_i) \quad\text{at }R=R_\text{x}\text{ },
\end{align}
\end{subequations}
where $\phi_i$ can be any channel-dependent phase that is constant in $R$ and energy \cite{PhysRevLett.81.3355}. The set of equations (\ref{eq:fhatandghat}) and their full radial derivatives define $\hat{f}$ and $\hat{g}$. As the WKB boundary conditions define these single-channel reference wave functions at the single radius $R_\text{x}$, they still constitute $\textit{exact}$ solutions of the radial Schr\"{o}dinger equation (\ref{eq:ref}).

Moreover, the particular solutions defined by (\ref{eq:fhatandghat}) have several advantages. First, unlike the wave functions (\ref{eq:bess}), these reference wave functions are not energy-normalized. They are well defined for $E_i\le0$ and analytic in energy across threshold. Second, as the large kinetic energy at short range dominates the low collision energies and relatively small Zeeman shifts of typical cold collisions, WKB boundary conditions lead to reference wave functions that are weakly dependent on collision energy and magnetic field. From this follows the weak energy and field dependence of $K^\text{sr}$. Third, this particular choice of boundary conditions allows $\hat{f}$ and $\hat{g}$ to be maximally linearly independent at short range.

The matrix $K^\text{sr}$ and the linearly independent reference wave functions $\hat{f}$ and $\hat{g}$ contain all the information necessary to calculate scattering observables. The quantum defect theory of \cite{PhysRevLett.81.3355} defines the four parameters $\eta$, $A$, $\mathcal{G}$, and $\beta_\text{Burke}$ that describe the asymptotic behavior of the wave functions $\hat{f}$ and $\hat{g}$. Hence, these parameters are also smooth functions of collision energy and magnetic field and completely describe the long-range physics. The notation in this paper only differs from the notation of \cite{PhysRevLett.81.3355} by the use of $\gamma$ instead of $\beta_\text{Burke}$, where $\cot\gamma=\tan\beta_\text{Burke}$. The introduction of $\gamma$ emphasizes its relationship with $\mathcal{G}$, and section \ref{sec:MQDT} demonstrates this relationship.

The calculation of $S^{\text{phys}}$ requires two linearly independent, energy-normalized wave functions at large $R$ in each open channel and the bound state wave function in each closed channel. To this end, the parameters $A$ and $\mathcal{G}$ create a Wronskian-preserving transformation between the reference wave functions $\hat{f}$ and $\hat{g}$ and two energy-normalized wave functions $f$ and $g$ for $E_i>0$, 
\begin{equation}
\label{eq:AandG}
\begin{pmatrix}
\ f \\
\ g \\
\end{pmatrix}
= 
\begin{pmatrix}
\ A^{1/2}& \ 0 \\
\ A^{-1/2} \mathcal{G} & A^{-1/2} \\
\end{pmatrix}
\begin{pmatrix}
\ \hat{f} \\
\ \hat{g} \\
\end{pmatrix} \text{ }.
\end{equation}
The parameter $A$ is responsible for the energy-normalization of $f$ and $g$, and the parameter $\mathcal{G}$ accounts for the different phase $\hat{g}$ accumulates in $V^\text{lr}$ than does $\hat{f}$. The phase shift $\eta$ describes how $f$ and $g$ differ from the spherical Bessel functions asymptotically,
\begin{subequations}
\label{eq:fandg}
\begin{align}
  \label{eq:f}
 f_i&\xrightarrow{R\rightarrow \infty} \frac{1}{\sqrt{k_i}}\sin(k_iR-L_i\pi/2+\eta_i)\text{ },\\
  \label{eq:g}
 g_i&\xrightarrow{R\rightarrow \infty} -\frac{1}{\sqrt{k_i}}\cos(k_iR-L_i\pi/2+\eta_i)\text{ }.
\end{align}
\end{subequations}
For $E_i<0$, the parameter $\gamma$ determines the linear combination of $\hat{f}$ and $\hat{g}$ that vanishes as $R\rightarrow\infty$,
\begin{equation}
 \label{eq:gamma}
 \tan\gamma_i\hat{f}_i+\hat{g}_i\xrightarrow{R\rightarrow\infty}\propto\ e^{-\kappa_iR}\text{ },
\end{equation}
where $\kappa_i=ik_i$. 

Calculating the four MQDT parameters requires the evaluation of several Wronskians that involve $\hat{f}$ and $\hat{g}$ as $R\rightarrow\infty$,
\begin{subequations}
\label{eq:MQDT}
\begin{equation}
 \tan\eta=\frac{W\left((kR)j_L(kR),\hat{f}\right)}{W\left((kR)n_L(kR),\hat{f}\right)}\Biggr|_{R\rightarrow\infty}\text{ },
\end{equation}\begin{equation}
 A^{-1}=\frac{W\left((kR)j_L(kR),\hat{f}\right)^2+W\left((kR)n_L(kR),\hat{f}\right)^2}{k}\Biggr|_{R\rightarrow\infty}\text{ },
\end{equation}\begin{equation}
\mathcal{G}=-\frac{W\left(g,\hat{g}\right)}{W\left(g,\hat{f}\right)}\Biggr|_{R\rightarrow\infty}\text{ },
\end{equation}\begin{equation}
\tan\gamma=-\frac{W\left(e^{-\kappa R},\hat{g}\right)}{W\left(e^{-\kappa R},\hat{f}\right)}\Biggr|_{R\rightarrow\infty}\text{ },
\end{equation}
\end{subequations}
where $W\left(y_1,y_2\right)$ is the Wronskian with respect to $R$ of any two functions $y_1$ and $y_2$,
\begin{equation}
 W\left(y_1,y_2\right)=y_1(R)\frac{\mathrm{d}y_2(R)}{\mathrm{d}R}-y_2(R)\frac{\mathrm{d}y_1(R)}{\mathrm{d}R}\text{ }.
\end{equation}

The MQDT parameters directly translate $K^{\text{sr}}$ into observables. By partitioning $K^\text{sr}$ into open (P) and closed (Q) channels, simple algebra produces the physical scattering matrix \cite{PhysRevLett.81.3355},
\begin{subequations}
\begin{equation}
 \label{eq:MQDTa}
 \tilde{K}=K_\text{PP}^\text{sr}-K_\text{PQ}^\text{sr}(K_\text{QQ}^\text{sr}+\cot\gamma)^{-1}K_\text{QP}^\text{sr}\text{ },
\end{equation}\begin{equation}
 K=A^{1/2}\tilde{K}(I+\mathcal{G}\tilde{K})^{-1}A^{1/2}\text{ },
\end{equation}\begin{equation}
 \label{eq:MQDTb}
 S^{\text{phys}}=e^{i\eta}\frac{I+iK}{I-iK}e^{i\eta}\text{ }.
\end{equation}
\end{subequations}

\section{\label{sec:standard}Standardizing MQDT}

Since the MQDT parameters clearly depend on the particular choice of reference wave functions, standardizing this choice allows the MQDT parameters for a particular long-range potential to be tabulated once and for all and defines a simple procedure to find $K^\text{sr}$ \cite{PhysRevLett.81.3355}. In general, the boundary conditions (\ref{eq:fhatandghat}) define an infinite family of reference wave functions -- one set for each value of $\phi_i$. This section identifies a value of $\phi_i$ that guarantees both the maximal numerical stability of the MQDT parameters as well as their smooth, analytic energy behavior. Moreover, since the calculation of $\phi_i$ is required to apply the boundary conditions (\ref{eq:fhatandghat}), this calculation must be numerically stable -- even at high $L$ -- to be useful.

\begin{figure}[t]
\subfigure[$\hat{f}$ asymptotically coincides with $\chi_+$ at zero energy.]{
\includegraphics[width=.99\columnwidth]{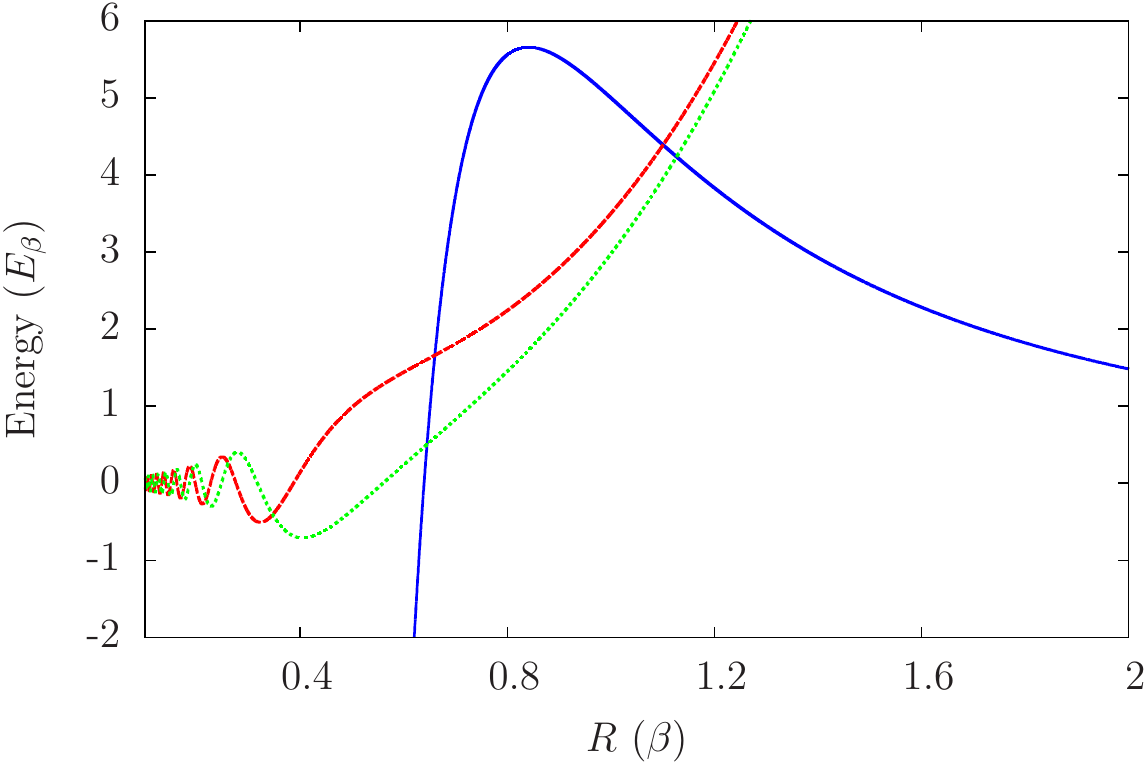}
\label{fig:oldref}
}
\subfigure[$\hat{g}$ asymptotically coincides with $\chi_-$ at zero energy]{
\includegraphics[width=.99\columnwidth]{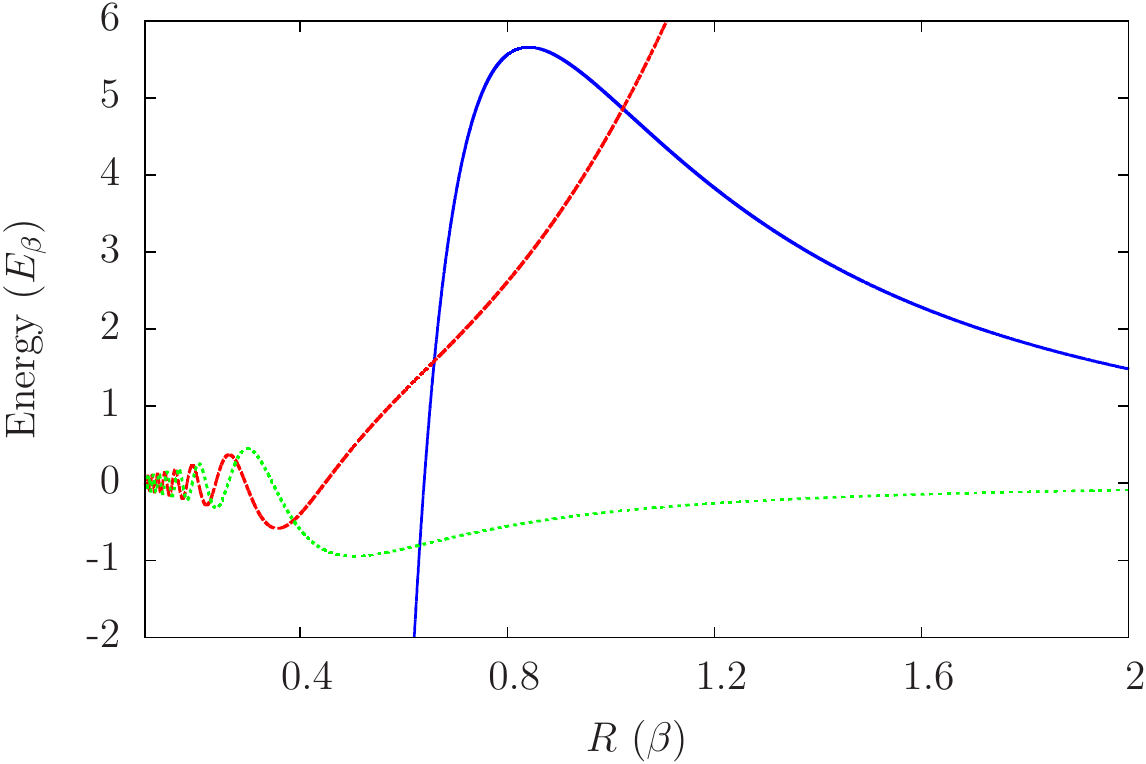}
\label{fig:newref}
}
\caption{(Color online) The reference wave functions $\hat{f}$ (red dashed curve) and $\hat{g}$ (green dotted curve) are shown for $E_i\approx 6.547\times 10^{-3}$ in the long-range potential $V^\text{lr}=-1/R^6$ (blue solid curve) with $L=2$. At short range, $\hat{f}$ and $\hat{g}$ are linearly independent. Under the classical barrier, these functions (a) lose their independence when $\hat{f}$ is chosen to asymptotically coincide with $\chi_+$ at zero energy and (b) retain their independence when $\hat{g}$ is chosen to asymptotically coincide with $\chi_-$ at zero energy. } 
\end{figure}

In the manner of \cite{PhysRevLett.81.3355}, the MQDT parameters are standardized by letting one of the reference wave functions asymptotically coincide (up to a normalization) with a particular wave function at $\textit{zero}$ energy. The choice of this standard, zero-energy wave function, therefore, identifies a particular value of $\phi_i$ and determines a particular $\hat{f}$ and $\hat{g}$ at zero energy. The energy dependence of $\hat{f}$ and $\hat{g}$ near threshold is then smoothly obtained from their WKB boundary conditions (\ref{eq:fhatandghat}). In principle, any value of $\phi_i$ is equally valid as long as $\hat{f}$ and $\hat{g}$ remain linearly independent. However, at ultracold energies, motion under a classical centrifugal barrier can often cause a pair of reference wave functions to become linearly dependent numerically, so it is our goal to find the value of $\phi_i$ that leads (in the limit of zero energy) to maximally linearly independent reference wave functions under and beyond this barrier. 

For all potentials that fall off faster than $1/R^2$, the asymptotic linear combination of $R^{L+1}$ and $R^{-L}$ uniquely identifies a particular zero-energy wave function. Consequently, the asymptotic linear combination of the analytic zero-energy solutions \cite{Oliver2010},
\begin{subequations}
\label{eq:zero}
\begin{align}
  \label{eq:chiplus}
 \chi_+&=\sqrt{R}J_{-\frac{1}{4}(2L+1)}(1/2R^2)\xrightarrow{R\rightarrow\infty}\propto R^{L+1}\text{ },\\
  \label{eq:chiminus}
 \chi_-&=\sqrt{R}J_{\frac{1}{4}(2L+1)}(1/2R^2)\xrightarrow{R\rightarrow\infty}\propto R^{-L}\text{ },
\end{align}
\end{subequations}
identifies a particular zero-energy wave function in any potential that is dominated by $V^\text{lr}=-1/R^6$ asymptotically. Here, $J$ is the Bessel function of the first kind. For $L>0$, these solutions take their asymptotic form under the classical barrier at $R\gtrsim1$. Therefore, all zero-energy wave functions start to resemble $\chi_+$ in this region -- losing their linear independence numerically -- except for those directly proportional to $\chi_-$.

Fig. \ref{fig:oldref} shows $\hat{f}$ and $\hat{g}$ in the long-range potential $V^\text{lr}=-1/R^6$ with $E_i\approx 6.547\times 10^{-3}$. This energy corresponds to $1$ $\mu$K for $^{40}$K + $^{87}$Rb when $C_6=4.300\times10^3$ in atomic units. Here, $\hat{f}$ is chosen to asymptotically coincide with $\chi_+$ at zero energy. The boundary conditions (\ref{eq:fhatandghat}) ensure the maximal independence of $\hat{f}$ and $\hat{g}$ at short range, but they both resemble $\chi_+$ in the classically forbidden region $R\gtrsim1$. While $\hat{f}$ is required to grow like $R^{L+1}$ under the classical barrier, $\hat{g}$ also quickly begins to grow in a similar way. Hence, this choice for $\hat{f}$ leads to a set of reference wave functions that exhibit increasing linear dependence as the collision energy approaches zero and the classical barrier grows. In fact, $\hat{g}$ always asymptotically diverges at zero energy except for a unique value of $\phi_i$. 

Since only $\chi_-$ remains numerically linearly independent from $\chi_+$ in the limit $R\rightarrow\infty$, letting $\hat{g}$ asymptotically coincide with $\chi_-$ at zero energy guarantees that the zero-energy limit of $\hat{g}$ is maximally independent from $\hat{f}$ not only at short range but also well into the classically forbidden region. Moving away from zero energy causes $\hat{g}$ to gain a contribution from $\chi_+$ asymptotically, but the classically forbidden region becomes smaller. Fig. \ref{fig:newref} shows that this choice for $\hat{g}$ leads to reference wave functions that are linearly independent at both short range and long range -- even at ultracold energies and high $L$. Hence, these reference wave functions are ideal for a numerical calculation of the MQDT parameters at ultracold energies.

In order to implement this standardization, one must determine the value of $\phi_i$ that is used to define $\hat{f}$ and $\hat{g}$. For long-range potentials in which the zero-energy solutions are known analytically at all $R$, this value of $\phi_i$ is easily derived using the values of $\chi_+$ and/or $\chi_-$ at $R_\text{x}$. However, since even the zero-energy solutions are only known analytically for a limited number of power law potentials, calculating $\phi_i$ numerically allows the use of the true long-range potential for a given scattering problem.

To this end the zero-energy reference wave functions for $\phi_i=0$ can be numerically propagated from their boundary conditions at $R_\text{x}<<1$ to large $R>>1$. At $R>>1$ the reference wave functions are well approximated by linear combinations of solutions that are known analytically, and $\tan\phi_i$ is a simple ratio of two Wronskians. For example, the value of $\tan\phi_i$ that lets $\hat{f}$ asymptotically coincide with $\chi_+$ at zero energy is given by,
\begin{subequations}
\begin{equation}
\label{eq:tanphi}
 \tan\phi_i\xrightarrow{R\rightarrow\infty}\frac{W\left(\chi_+,\hat{f}_{(\phi_i=0)}\right)}{W\left(\chi_+,\hat{g}_{(\phi_i=0)}\right)}\text{ }.
\end{equation}
An alternative choice allows $\hat{g}$ to asymptotically coincide with $\chi_-$ at zero energy, and this value of $\phi_i$ is given by,
\begin{equation}
\label{eq:tanphi2}
 \tan\phi_i\xrightarrow{R\rightarrow\infty}-\frac{W\left(\chi_-,\hat{g}_{(\phi_i=0)}\right)}{W\left(\chi_-,\hat{f}_{(\phi_i=0)}\right)}\text{ }.
\end{equation}
\end{subequations}

\begin{figure}[t]
\subfigure[$L=0$]{
    \label{fig:subfig1}
    \includegraphics[width=.46\columnwidth]{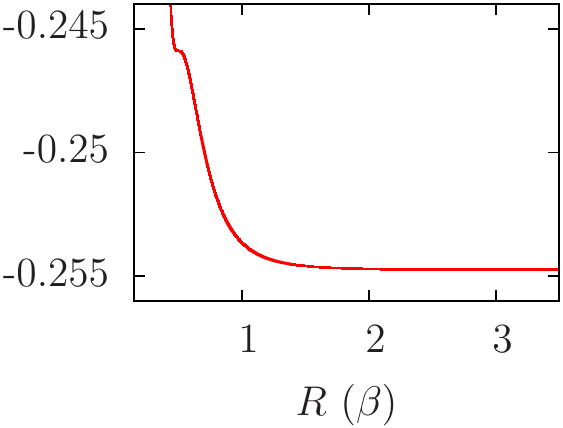}
}
\subfigure[$L=1$]{
    \label{fig:subfig2}
    \includegraphics[width=.46\columnwidth]{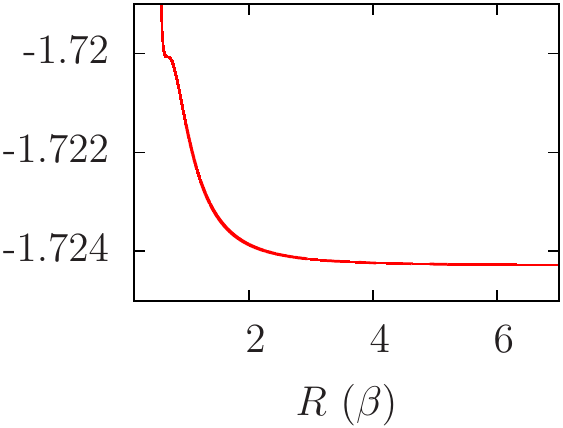}
}
\subfigure[$L=2$]{
    \label{fig:subfig3}
    \includegraphics[width=.46\columnwidth]{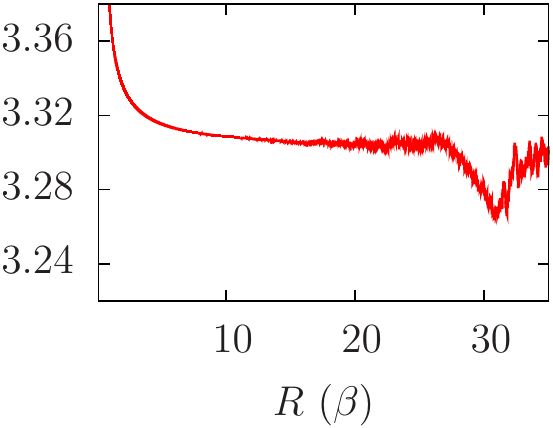}
}
\subfigure[$L=3$]{
    \label{fig:subfig4}
    \includegraphics[width=.46\columnwidth]{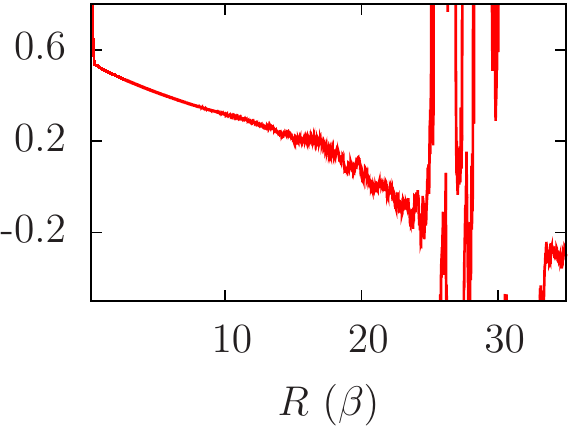}
}
\caption[]{The convergence of $\tan\phi_i$ with $R$ is shown when $L=0-3$. Here, $\hat{f}$ is chosen to asymptotically coincide with $\chi_+$ at zero energy in the long-range potential $V^\text{lr}=-C_6/R^6-C_8/R^8-C_{10}/R^{10}$, and $R_\text{x}=.1$.}
\label{fig:tanz}
\end{figure}
\begin{figure}[t]
\subfigure[$L=0$]{
    \label{fig:subfig1}
    \includegraphics[width=.46\columnwidth]{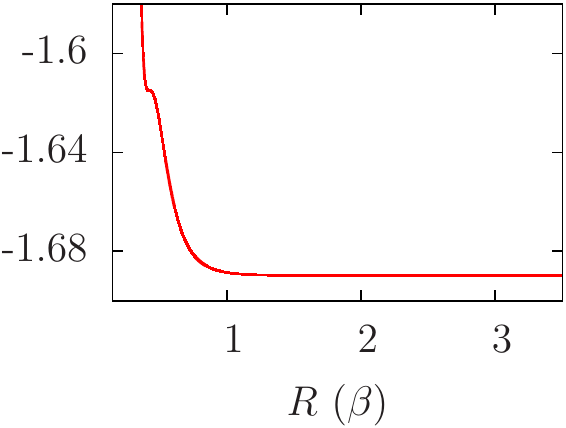}
}
\subfigure[$L=2$]{
    \label{fig:subfig3}
    \includegraphics[width=.46\columnwidth]{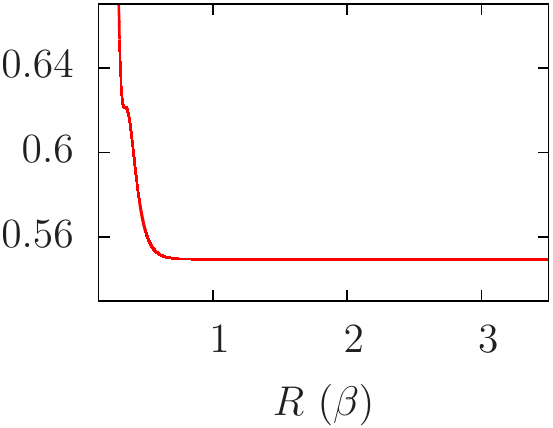}
}
\subfigure[$L=10$]{
    \label{fig:subfig5}
    \includegraphics[width=.46\columnwidth]{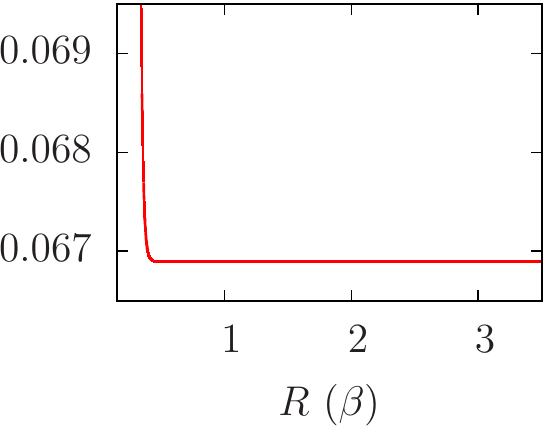}
}
\subfigure[$L=20$]{
    \label{fig:subfig5}
    \includegraphics[width=.46\columnwidth]{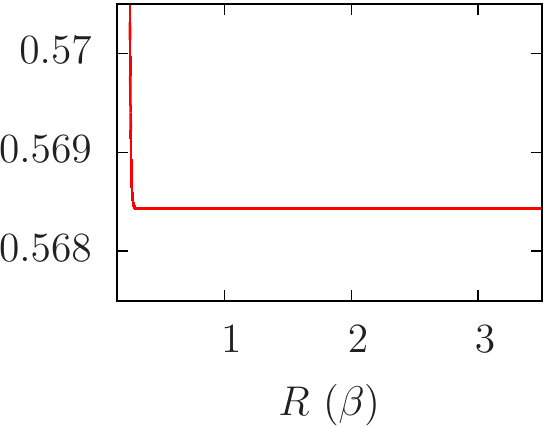}
}
\caption[]{The convergence of $\tan\phi_i$ with $R$ is shown when $L=0$, $2$, $10$, and $20$. Here, $\hat{g}$ is chosen to asymptotically coincide with $\chi_-$ at zero energy in the long-range potential $V^\text{lr}=-C_6/R^6-C_8/R^8-C_{10}/R^{10}$, and $R_\text{x}=.1$.}
\label{fig:tany}
\end{figure}

The stable calculation of $\phi_i$ requires the linear independence of the reference wave functions $\hat{f}_{(\phi_i=0)}$ and $\hat{g}_{(\phi_i=0)}$. As they each represent a particular linear combination of $\chi_+$ and $\chi_-$ asymptotically, their linear independence relies on their different, and quickly vanishing, contributions from $\chi_-$. Hence, they can easily lose their linear independence as $R$ grows, causing the calculation of $\phi_i$ to become unstable. For example, Fig. \ref{fig:tanz} demonstrates the difficulty of numerically calculating $\phi_i$ by showing the numerically computed ratio of Wronskians in (\ref{eq:tanphi}) for all values of $R$. These calculations use the long-range potential $V^\text{lr}=-C_6/R^6-C_8/R^8-C_{10}/R^{10}$ for several values of $L$, where the dispersion coefficients are $C_6=4.300\times 10^3$, $C_8=4.823\times 10^5$, and $C_{10}=6.181\times 10^7$ in atomic units. These values are realistic for collisions of K + Rb \cite{pashov:022511}.

Fig. \ref{fig:tanz} shows that the calculation of $\tan\phi_i$, at least for $L=0$, converges quickly after $R=1$, where the $-C_8/R^8$ and $-C_{10}/R^{10}$ terms of the long-range potential become dominated by the $-C_6/R^6$ term. However, as $L$ increases, a larger value of $R$ is required to converge this calculation, and Fig. \ref{fig:tanz} shows that a converged calculation of $\tan\phi_i$ is not numerically stable for $L>1$. Appendix \ref{app:tanz} explores this instability and determines that, for $L>1$, the asymptotic contribution to each reference wave function from $\chi_-$ is eventually dominated by the presence of finite numerical noise, causing the calculation of $\tan\phi_i$ to be unstable. If even $\phi_i$ cannot be stably computed, calculating the MQDT parameters is hopeless.

However, Appendix \ref{app:tanz} also shows that this instability vanishes if $\hat{g}$ is chosen to asymptotically coincide with $\chi_-$ at zero energy -- the same choice as in Fig. \ref{fig:newref}. For $\textit{only}$ this choice of $\hat{g}$, the calculation of $\phi_i$ does not require finding the asymptotic contribution to $\hat{f}_{(\phi_i=0)}$ and $\hat{g}_{(\phi_i=0)}$ from $\chi_-$. Hence, in this case, the calculation of $\tan\phi_i$ is numerically stable for all $L$. In contrast to Fig. \ref{fig:tanz}, Fig. \ref{fig:tany} demonstrates that $\tan\phi_i$ stably converges with $R$ by showing the numerically computed ratio of Wronskians in (\ref{eq:tanphi2}) for all values of $R$. These calculations use the same long-range potential as the calculations in Fig. \ref{fig:tanz} and include much larger values of $L$. 

The calculations of $\tan\phi_i$ in Fig. \ref{fig:tany} converge more rapidly as $L$ increases. This trend is intuitive because, as $L$ grows, the classical turning point at zero energy moves inward and $\chi_-$ becomes increasingly distinct from all other solutions. Moreover, these calculations of $\tan\phi_i$ are stable out to very large $R\gtrsim10^3$. This allows for a more consistent calculation of the MQDT parameters because the value of $R$ at which $\phi_i$ is actually determined can equal the value of $R$ that is necessary for a well-converged calculation of the MQDT parameters. Although either $\hat{f}$ or $\hat{g}$ could asymptotically coincide with $\chi_-$ at zero energy, choosing $\hat{g}$ for this role defines our standardization because the MQDT parameters acquire very appealing qualities. The next section explores these qualities.

\section{\label{sec:MQDT}Calculating MQDT parameters}

\begin{figure}[t]
\includegraphics[width=.99\columnwidth]{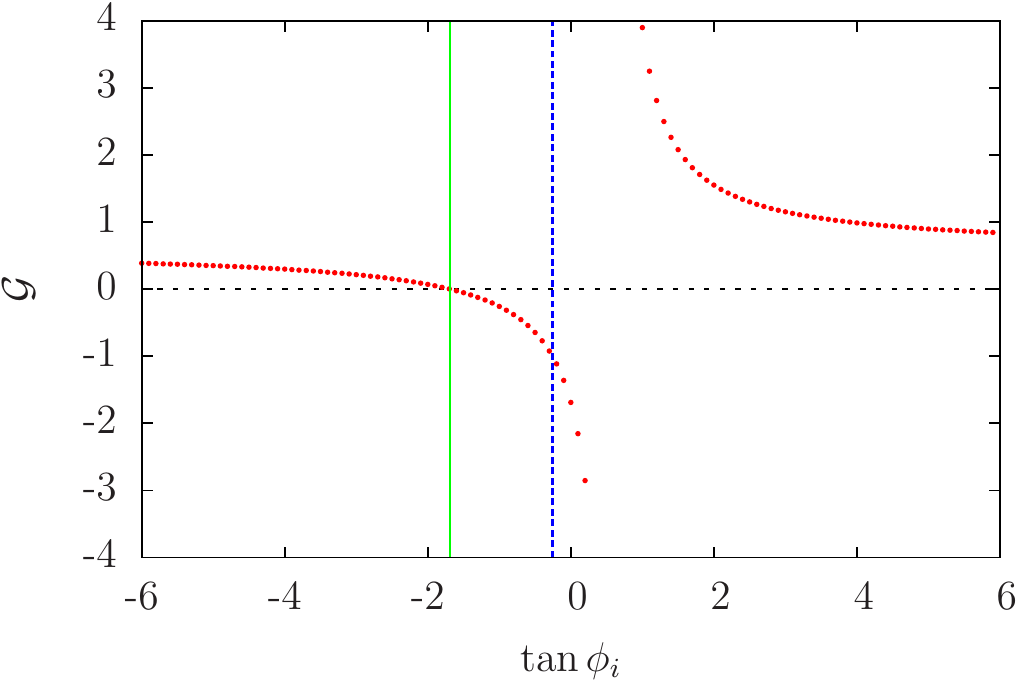}
\caption{\label{fig:Gvphi} (Color online) The red dots are the MQDT parameter $\mathcal{G}$ shown as a function of $\tan\phi_i$. The vertical lines represent the value of $\tan\phi_i$ when (blue dashed line) $\hat{f}$ asymptotically coincides with $\chi_+$ at zero energy and (green solid line) $\hat{g}$ asymptotically coincides with $\chi_-$ at zero energy. For this calculation, $V^\text{lr}=-C_6/R^6-C_8/R^8-C_{10}/R^{10}$, where the dispersion coefficients are appropriate for K + Rb; $E_i\approx 6.547\times 10^{-4}$, which corresponds to $100$ nK; $L=0$; and $R_\text{x}=.1$.} 
\end{figure}

\begin{figure*}[ht]
\subfigure[$L=0$]{
\includegraphics[width=.65\columnwidth]{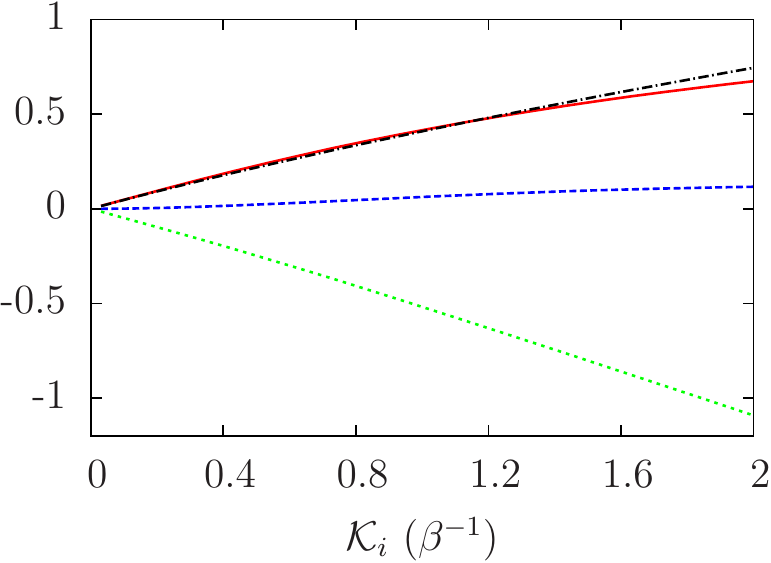}
}
\subfigure[$L=1$]{
\includegraphics[width=.65\columnwidth]{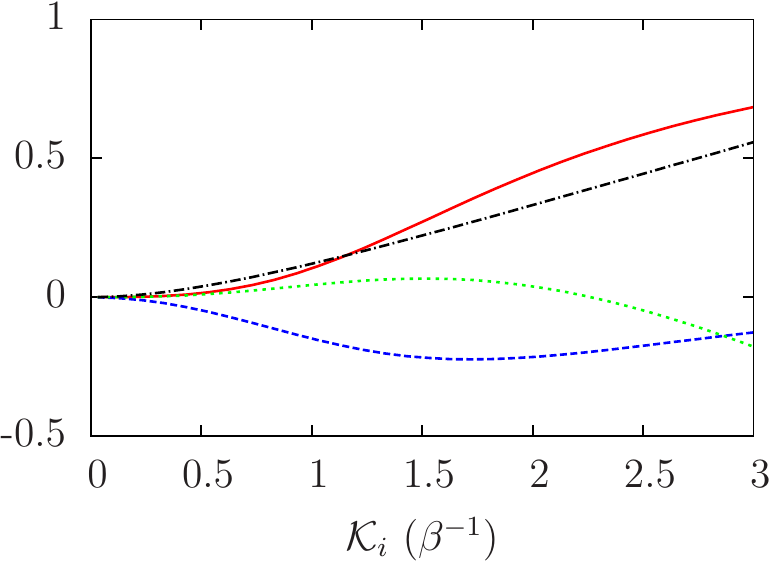}
}
\subfigure[$L=2$]{
\includegraphics[width=.65\columnwidth]{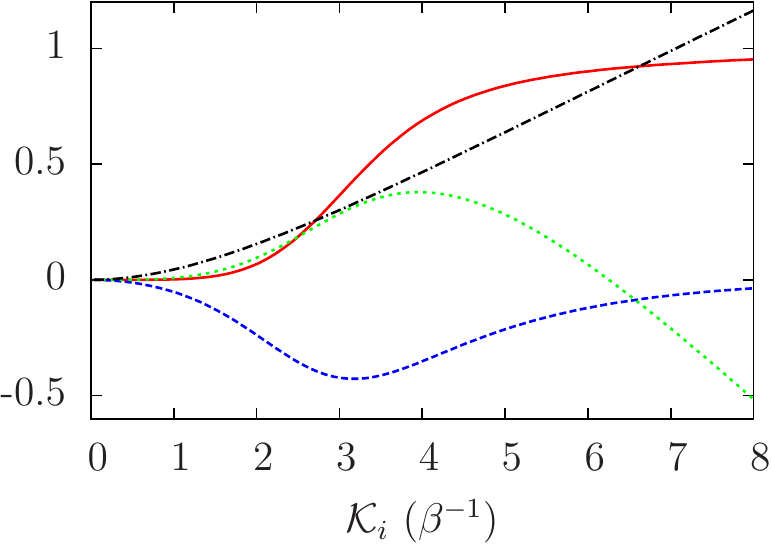}
}
\subfigure[$L=3$]{
\includegraphics[width=.65\columnwidth]{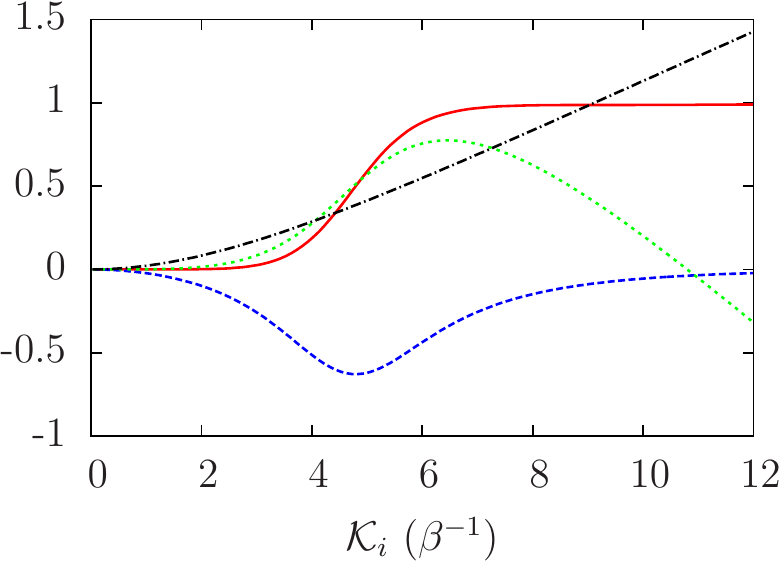}
}
\subfigure[$L=4$]{
\includegraphics[width=.65\columnwidth]{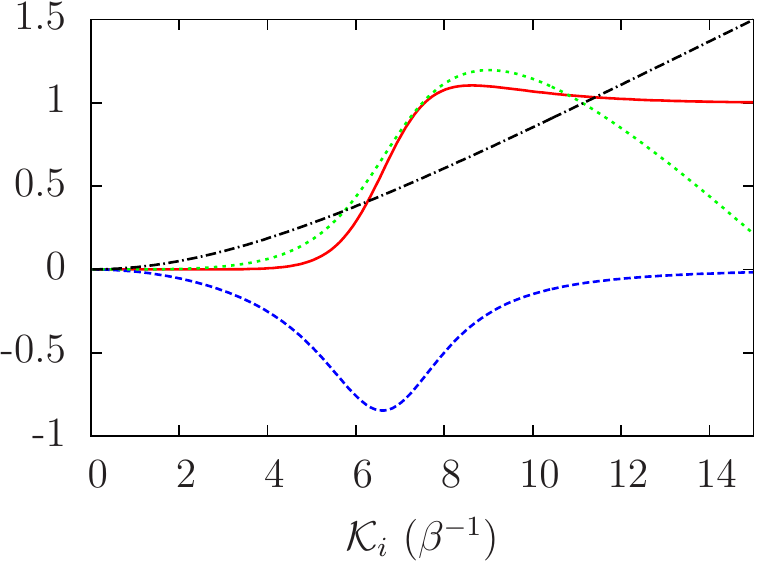}
}
\subfigure[$L=5$]{
\includegraphics[width=.65\columnwidth]{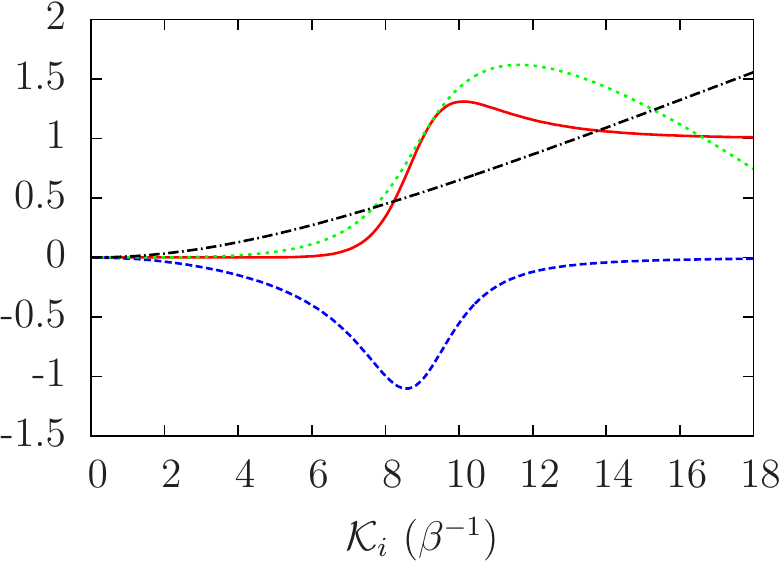}
}
\caption{\label{fig:qdt03} (Color online) The MQDT parameters $A$ (red solid curve), $\mathcal{G}$ (blue dashed curve), $\eta$ (green dotted curve), and $\gamma$ (black dashed-dotted curve) are shown for the KRb model potential $V^\text{lr}=-C_6/R^6-C_8/R^8-C_{10}/R^{10}$ with $L=0-5$. The parameter $\mathcal{K}_i\equiv\sqrt{|E_i|}$ is defined merely for plotting purposes.}
\end{figure*}

Having specified $\phi_i$ in (\ref{eq:fhatandghat}), the values of the MQDT parameters $A$, $\eta$, $\mathcal{G}$, and $\gamma$ follow unambiguously from (\ref{eq:MQDT}). Of particular importance is the parameter $\mathcal{G}$. The zero-energy limit of $\mathcal{G}$,
\begin{equation}
 \label{eq:Gzero}
 \mathcal{G}=-\frac{W\left(g,\hat{g}\right)}{W\left(g,\hat{f}\right)}\Biggr|_{R\rightarrow\infty}\xrightarrow{E_i\rightarrow0} -\frac{W\left(\chi_-,\hat{g}\right)}{W\left(\chi_-,\hat{f}\right)}\Biggr|_{R\rightarrow\infty}\text{ },
\end{equation}
is intimately related to the value of $\phi_i$ in (\ref{eq:tanphi2}). Clearly, if $\phi_i=0$ in (\ref{eq:Gzero}), the zero-energy limit of $\mathcal{G}$ is the exact same value as $\tan\phi_i$ in (\ref{eq:tanphi2}). Appendix \ref{app:tanz} shows that evaluating (\ref{eq:tanphi2}) is stable independent of the phase $\phi_i$ given to the reference wave functions. Therefore, the zero-energy limit of $\mathcal{G}$ is numerically stable. Fig. \ref{fig:Gvphi} shows the calculation of $\mathcal{G}$ at $100$ nK as a function of the zero-energy phase $\phi_i$.

For a given $R_\text{x}$, Fig. \ref{fig:Gvphi} shows that there is a unique value of $\phi_i$ (indicated by the green vertical solid line) for which $\mathcal{G}$ vanishes at zero energy. This is the choice where $\hat{g}$ asymptotically coincides with $\chi_-$ at zero energy; therefore, $\mathcal{G}=0$ at zero energy defines our standardization. As evident from (\ref{eq:AandG}) and (\ref{eq:fandg}), the vanishing of $\mathcal{G}$ guarantees the maximal linear independence of $\hat{f}$ and $\hat{g}$ at long range. Hence, although a different standardization could guarantee that the zero-energy limit of $\mathcal{G}$ is well behaved (e.g., the blue vertical dashed line in Fig. \ref{fig:Gvphi}), our choice of reference wave functions is the $\textit{only}$ choice that is maximally linearly independent at long range in the limit of zero energy.
 
Applying our standardization, Fig. \ref{fig:qdt03} illustrates the energy dependence of the various MQDT parameters for the KRb model potential $V^\text{lr}=-C_6/R^6-C_8/R^8-C_{10}/R^{10}$, where the dispersion coefficients are the same as in the previous section. The positive and negative energy parameters are plotted together by defining the parameter $\mathcal{K}_i\equiv\sqrt{|E_i|}$. The MQDT parameters are presented in the natural van der Waals units $\beta=(2 \mu C_6 / \hbar^2)^{1/4}=143.9$ a$_0$ and $E_\beta=152.7$  $\mu$K. Each panel represents the result for a different partial wave $L$; note that a greater energy range is shown for higher $L$. In all cases the calculation is numerically stable, even in the threshold limit. These functions are smooth and hence easily interpolated.

One striking feature, unique to our parameterization, is that all parameters vanish as powers of $E_i$ in the $E_i\rightarrow0$ limit. In this limit they are well approximated by simple analytic formulas. For alkali atoms, where $C_8/R^8$ and $C_{10}/R^{10}$ make small corrections to $C_6/R^6$, these formulas can be derived using the $-C_6/R^6$ potential alone.  Their derivation is detailed in Appendix \ref{app:thresh}, and the results are summarized below. These parameters are conveniently parameterized in terms of a set of generalized, standard scattering lengths $\bar{a}_L$,
\begin{equation}
 \label{eq:abar}
 \bar{a}_L=\left(\frac{\pi 2^{-(2L+3/2)}}{\Gamma(L/2+5/4)\Gamma(L+1/2)}\right)^{2/(2L+1)}\text{ }.
\end{equation}
The threshold behavior of all four MQDT parameters is given here,
\begin{subequations}
\label{eq:thresh}
\begin{align}
 A^{1/2}&\xrightarrow{k\rightarrow0}-(\bar{a}_Lk)^{L+1/2}\text{ },\\
 \label{eq:phaseshift}
 \eta&\xrightarrow{k\rightarrow0}(-1)^{L+1}(\bar{a}_Lk)^{2L+1}+\frac{3 \pi \Gamma(L-3/2)}{32\Gamma(L+7/2)}k^4\text{ },\\
 \mathcal{G}&\xrightarrow{k\rightarrow0}(-1)^{L+1}(\bar{a}_Lk)^{4L+2}-\frac{k^2}{(2L+3)(2L-1)}\text{ },\\
 \gamma&\xrightarrow{\kappa\rightarrow0}
 \begin{cases}
 \bar{a}_0\kappa&\text{for }L=0\text{ },\\
 \frac{\kappa^2}{(2L+3)(2L-1)}&\text{for }L>0\text{ }.
\end{cases}
\end{align}
\end{subequations}
These formulas agree well with the numerical results for the exact long-range potential when $E_i\lesssim1$. 

Expressions such as (\ref{eq:abar}) have been derived before in the literature. For example, $\bar{a}_0$, which is the scattering length of our reference wave function $\hat{f}$, coincides with the semiclassical scattering length of Gribakin and Flambaum \cite{PhysRevA.48.546} for the $-C_6/R^6$ potential. Likewise, using an exact solution that was expressed using continued fractions \cite{PhysRevA.58.1728} \cite{Gao2004} \cite{PhysRevA.80.012702}, Gao performed a similar analytic treatment of the near-threshold MQDT parameters for the $-C_6/R^6$ potential. To do so, he identified a set of standard constants $\bar{a}_{sL,\text{Gao}}$ that are related to our equation (\ref{eq:abar}) via $\bar{a}_{sL,\text{Gao}} = (\bar{a}_L)^{2L+1}$.  While the treatments are equivalent, our parameters $\bar{a}_L$ have units of length. Moreover, our standard and universal reference wave functions have a universal form (\ref{eq:phaseshift}) for the corresponding phase shift.

Gao conceives of a hierarchy of reference wave functions distinguished by a short-range quantum defect parameter $\mu^c$ \cite{gao:012702}. Our implementation of MQDT introduces alternative short-range phases $\phi_i$ in (\ref{eq:fhatandghat}). Our particular choice for $\phi_i$ gives our reference wave function $\hat{f}$ a particular set of scattering lengths $\bar{a}_L$. Using equation (10) of \cite{PhysRevA.80.012702} and the zero-energy limit of equation (33) of \cite{PhysRevA.80.012702}, this choice corresponds to,
\begin{equation}
 \label{eq:Gao}
 \mu^c=1/2+L/4\text{ }.
\end{equation}
Equation (\ref{eq:Gao}) gives an explicit connection between the analytic formulas of Gao and our formulation of MQDT for the pure $-C_6/R^6$ potential.

Moreover, a similar formulation of MQDT \cite{PhysRevLett.104.113202} has derived the threshold behavior of three positive energy parameters for a reference potential with arbitrary scattering length $a$. These parameters are easily related to our MQDT parameters for the special case of $L=0$. Expressed in our notation, the results of \cite{PhysRevLett.104.113202} read,
\begin{align}
 \eta_{(L=0)}&\xrightarrow{k\rightarrow0}-ak\text{ },\\
 A_{(L=0)}&\xrightarrow{k\rightarrow0}\bar{a}_0k\left(1 + (a/\bar{a}_0 - 1)^2\right)\text{ },\\
 \mathcal{G}_{(L=0)}&\xrightarrow{k\rightarrow0}1-a/\bar{a}_0\text{ }.
\end{align}
If $a=\bar{a}_0$ here, these formulas are consistent with the threshold behavior of our set of MQDT parameters (\ref{eq:thresh}).

Although Appendix \ref{app:thresh} explicitly derives the MQDT parameter threshold behavior for only the $-1/R^6$ potential, this analysis implies simple extensions of (\ref{eq:thresh}) for any $-1/R^n$ potential. Here, $n$ is any integer ($n>2$). The zero-energy solution $\chi_-$ is well defined (and known analytically) for any potential of this kind,
\begin{equation}
 \chi_-=\sqrt{R}J_\nu\left(\frac{R^{-(2L+1)/2\nu}}{(2L+1)/2\nu}\right)\xrightarrow{R\rightarrow\infty}\propto R^{-L}\text{ },
\end{equation}
where $\nu=(2L+1)/(n-2)$. Hence, for any potential asymptotically dominated by $-C_n/R^n$ and for all $L$, our standardization uniquely specifies the zero-energy limit of $\hat{f}$ and $\hat{g}$, and the analysis of Appendix \ref{app:thresh} is repeatable. As our standardization demands that all MQDT parameters go to zero in the limit $E_i\rightarrow0$, $\hat{f}$ and $\hat{g}$ have maximal linear independence at zero energy -- even for this more general potential. A future report will consider the full theory for arbitrary $n$ and $L$.



\section{\label{sec:KRb} High Partial Wave Resonances}

Performing a full, numerical calculation of scattering observables allows the results of our formulation of MQDT to be tested. Here, the Johnson log-derivative propagator method \cite{Johnson1973445} produces numerically exact solutions to the coupled Schr\"{o}dinger equations ($\ref{eq:SE}$). This method propagates the log-derivative matrix $Y$ to very long range $R\approx20$, where it approaches a constant. After this full close-coupling (FCC) propagation, the asymptotic limit of $Y$ determines all scattering observables. The FCC calculation is known to be quite accurate, and it is our standard with which to compare the accuracy of MQDT.

The prediction of high-$L$ resonances in atomic collisions often requires many ($N>>1$) channels, causing the FCC calculations to become very time-consuming. The time required to compute the scattering matrix at a single energy and magnetic field is proportional to $N^3$. Moreover, resonance widths decrease quickly with $L$. Thus, the prediction of resonance locations and widths requires many long FCC calculations. However, if the elements of $K^\text{sr}$ are weakly dependent on energy and magnetic field, MQDT describes the same resonances after only a few iterations of a fraction of the full calculation. For instance, interpolating $K^{\text{sr}}$ on a coarse grid in magnetic field greatly increases the numerical efficiency of calculating Fano-Feshbach resonances.

\begin{figure}
\includegraphics[width=.99\columnwidth]{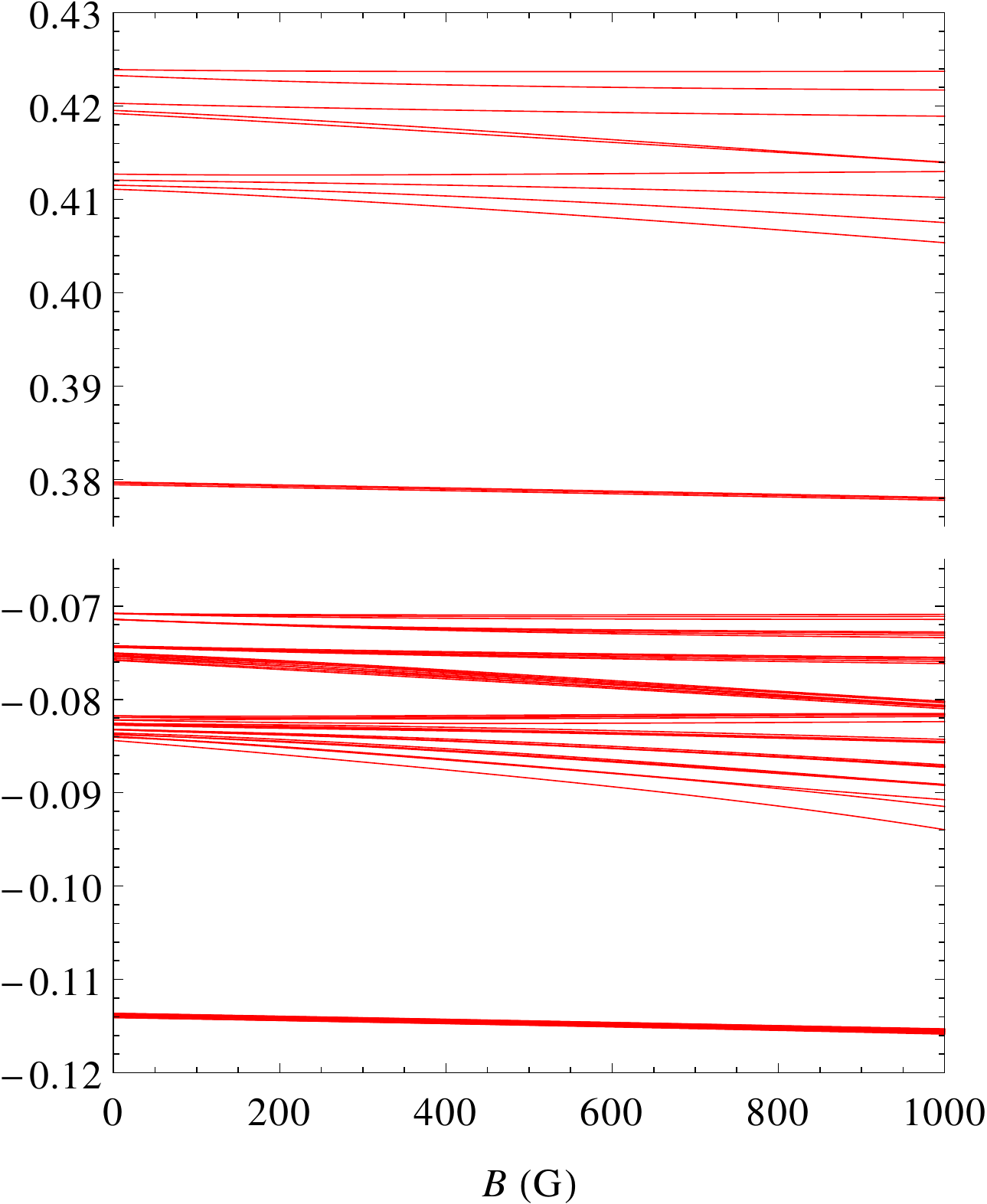}
\caption{\label{fig:Ksr} The eigenphase shifts $\mu_\lambda$ are shown for the collision of $^{40}$K + $^{87}$Rb over the range $B=0-1000$ G with a collision energy of $1$ $\mu$K. For this calculation the log-derivative matrix $Y$ is matched to solutions in the long-range potential $V^\text{lr}=-C_6/R^6-C_8/R^8-C_{10}/R^{10}$ at $R_\text{m}=45$ a$_0$, including channels with $L=0$ and $L=2$. The two graphs merely display two different ranges of $\mu$ over the same range of $B$.} 
\end{figure}

Consider, for example, the collision of $^{40}$K + $^{87}$Rb in the lowest hyperfine states $\ket{F_K ,M_{F_K}}\ket{F_{Rb},M_{F_{Rb}}}=\ket{9/2, -9/2}\ket{1,1}$, where a number of Fano-Feshbach resonances have been observed. Using the same model potential for both the FCC calculation and the calculation of $K^\text{sr}$ allows for a direct comparison between the FCC and MQDT methods. Our model adopts the accurate, short-range $X$ $^1\Sigma^+$ and $a$ $^3\Sigma^+$ molecular potentials of \cite{pashov:022511} that were constructed by performing a global fit to the position of the $L=0$ Fano-Feshbach resonances. For consistency, our model also adopts the interaction parameters of \cite{pashov:022511} that describe the long-range forces. The van der Waals parameters $C_6$, $C_8$, and $C_{10}$ describe the long-range dispersion forces; the electron-exchange interaction is,
\begin{equation}
 E_{\text{ex}}=A_{\text{ex}}R^{\gamma_{\text{ex}}}e^{-\beta_{\text{ex}} R}\text{ },
\end{equation}
which is added to the triplet molecular state and subtracted from the singlet molecular state; and the dipole-dipole interaction is,
\begin{equation}
 H_{\text{dd}}=-\frac{\alpha^2}{2}\left(3S_z^2-S^2\right)\left(1/R^3+a_{\text{SO}}e^{-b_{\text{SO}}(R-R_{\text{SO}})}\right)\text{ },
\end{equation}
where $\alpha$ is the fine structure constant.

Beyond $R_\text{m}$, the MQDT and FCC calculations involve slightly different Hamiltonians. On the one hand, the MQDT reference wave functions are solutions in the long-range potential of our choice. Because the anisotropic dispersion forces dominate at long range, our choice is
\begin{equation}
 V^{lr}=-C_6/R^6-C_8/R^8-C_{10}/R^{10}\text{ }.
\end{equation}
As a consequence, each channel differs only by a constant energy set by its hyperfine quantum numbers and subsequent Zeeman shift. This choice of $V^{\text{lr}}$ ignores all other forces and all couplings between channels beyond $R_\text{m}$. On the other hand, the FCC calculation considers the full Hamiltonian into the asymptotic region.

However, $E_{\text{ex}}$ is vanishingly small beyond $R\approx30$ a$_0$; hence, only $H_\text{dd}$ is responsible for the difference between the MQDT and FCC calculations. $H_\text{dd}$ is very long range and creates a coupling between channels, but its inclusion beyond $R_\text{m}=45$ a$_0$ makes a negligible contribution to the elastic cross section in this case. This allows for excellent agreement between the two calculations. In applications where such weak longer-range couplings must be included, it should be straightforward to include them perturbatively, along the lines formulated, for instance, in \cite{Badnell1999} \cite{Gorczyca2000} \cite{PhysRevA.63.032714}.

If one chooses an $R_\text{m}$ where all channels are locally open, typically $R_\text{m}\le35-50$ a$_0$ for alkali atoms, all resonant behavior is due to physics beyond this range and, therefore, approximately described by the MQDT parameters. Hence, the choice of $R_\text{m}=45$ a$_0$ leads to a smooth $K_\text{sr}$ that is easy to interpolate over a large range of collision energy and magnetic field. The eigenvalues of $K_\text{sr}$ are known as eigenphase shifts $\mu_{\lambda}=\tan\delta_\lambda/\pi$. Fig. \ref{fig:Ksr} shows these eigenphase shifts as a function of magnetic field over the range $B=0-1000$ G.

With this same choice of $R_\text{m}=45$ a$_0$, a very coarse magnetic field grid of spacing $100$ G allows for an accurate interpolation of $K^\text{sr}$, and MQDT accurately reproduces the measured Fano-Feshbach resonances in $L=0-2$ states \cite{simoni:052705}. This accuracy and the excellent agreement with the FCC calculation motivate using MQDT to quickly re-fit the singlet and triplet scattering lengths, producing our own accurate scattering model. Our fit includes the experimentally measured resonance positions of \cite{simoni:052705} in all $L\le2$ states. It also includes the $L=2$ resonance at $547.4(1)$ G reported in \cite{PhysRevA.74.041605} and recently confirmed by \cite{RuthKRb}. While retaining the value of $C_6=4.300\times 10^3$ atomic units, varying the scattering lengths leads to a minimum reduced chi-squared between the experimental and MQDT resonance positions. The optimal scattering lengths are $a_\text{s}=-110.8$ a$_0$ and $a_\text{t}=-214.5$ a$_0$ with $\chi^2_{\text{red}}=0.83$.

\begin{figure}
\includegraphics[width=.99\columnwidth]{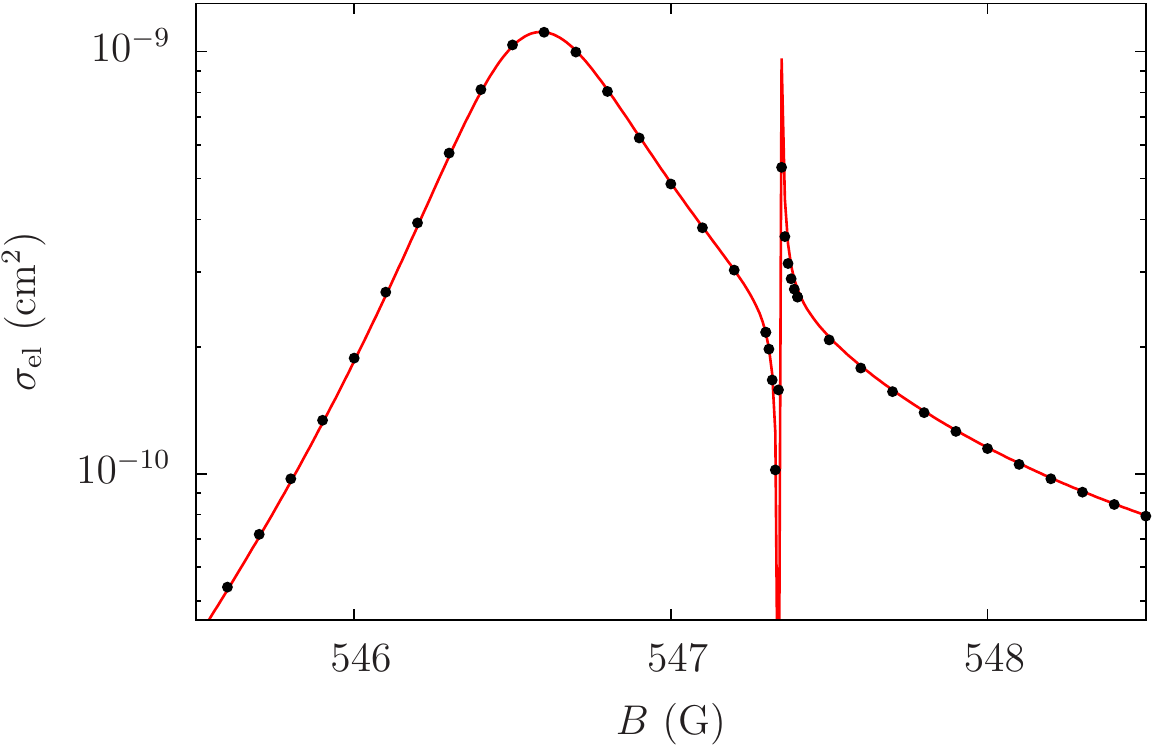}
\caption{\label{fig:RuthRes} (Color online) The elastic cross section for the collision of $^{40}$K + $^{87}$Rb with a collision energy of $1$ $\mu$K is shown for the FCC calculation (black dots), including channels with $L=0$ and $L=2$. The FCC calculation is compared to the MQDT calculation (red curve) with $K_\text{sr}$ interpolated over a range of $1000$ G. This curve is unchanged if the analytic formulas (\ref{eq:thresh}) are used instead of the numerical values for $A$, $\eta$, and $\mathcal{G}$.} 
\end{figure}

Using this re-tuned Hamiltonian, our model predicts the position and width of the $L=0-2$ Fano-Feshbach resonances for $^{40}$K + $^{87}$Rb collisions in their lowest hyperfine states. For example, Fig. \ref{fig:RuthRes} shows the MQDT and FCC calculations for overlapping s- and d-wave resonances. By only calculating $K_\text{sr}$ once every $100$ G and interpolating over the range $B=0-1000$ G, MQDT reproduces the FCC calculation of resonance positions with an accuracy of $\lesssim1$ mG. Moreover, since using MQDT to search for resonances only requires a magnetic field grid finer than the distance between any two resonances \cite{Suleimanov2011}, the method allows enough numerical efficiency to ensure the discovery of $\textit{all}$ the Fano-Feshbach resonances in this range of magnetic field. 

Table \ref{tab:reslista} lists these resonance positions and widths. Even though some resonances are very narrow, finding the roots of det$\left(K^\text{sr}_\text{QQ}+\cot\gamma\right)$ determines quantitatively accurate resonance positions \cite{PhysRevLett.81.3355}, where both $K^\text{sr}$ and $\gamma$ are interpolated with ease. As our model calculates all experimentally measured resonances close to their positions $B_\text{ex}$, the unmeasured resonance positions of Table \ref{tab:reslista} are predictive with uncertainties on the order of current experimental uncertainties ($\lesssim 1$ G). Once our theory predicts a resonance at the position $B_\text{th}$, fitting the divergence of the scattering length near the resonance to the following form determines the width of an $L=0$ or $L=2$ resonance \cite{simoni:052705},
\begin{equation}
  \label{eq:width}
 a(B) = a_\text{bg}\left(1-\frac{\Delta}{B-B_\text{th}}\right)\text{ },
\end{equation}
where $a_\text{bg}$ is the local background scattering length and $\Delta$ is the field width. Fitting the divergence of the scattering volume to the same form as equation (\ref{eq:width}) determines the width of an $L=1$ resonance. 


\begin{table}
\caption{\label{tab:reslista} All of the Fano-Feshbach resonances in the range $B=0-1000$ G for the collision of $^{40}$K + $^{87}$Rb in the state $\ket{F_K ,MF_K}\ket{F_{Rb},MF_{Rb}}=\ket{-9/2 -9/2}\ket{1,1}$ are calculated using MQDT for a collision energy of $1$ $\mu$K. These resonance positions $B_{\text{th}}$ and field widths $\Delta$ are listed here with their associated partial wave quantum number $L$ and compared with experimentally measured resonance positions $B_{\text{ex}}$. Note: All magnetic field values are in units of Gauss.
}
\begin{ruledtabular}
\begin{tabular}{cccc||cccc}
$B_{\text{ex}}$ & $B_{\text{th}}$ & $-\Delta$ & $L$ & $B_{\text{ex}}$ & $B_{\text{th}}$ & $-\Delta$ & $L$ \\
\hline
- & 96.06 & $1.5\times 10^{-16}$ & 2 & - & 506.3 & $4.5\times 10^{-6}$ & 2\\
- & 108.7 & $2.3\times 10^{-13}$ & 2 & 515.7(5) & 515.1 & $0.50$ & 1\\
- & 124.3 & $3.1\times 10^{-9}$ & 2 & - & 526.5 & $5.7\times 10^{-5}$ & 1\\
- & 143.9 & $2.6\times 10^{-8}$ & 2 & - & 531.8 & $3.1\times 10^{-5}$ & 1\\
- & 155.9 & $1.9\times 10^{-12}$ & 1 & - & 540.8 & $1.9\times 10^{-6}$ & 1\\
- & 168.1 & $2.7\times 10^{-7}$ & 2 & 546.6(2) & 546.6 & $3.1$ & 0\\
- & 171.9 & $1.1\times 10^{-11}$ & 2 & 547.4(1) & 547.3 & $6.3\times 10^{-3}$ & 2\\
- & 178.2 & $3.7\times 10^{-6}$ & 1 & - & 558.5 & $9.6\times 10^{-8}$ & 2\\
- & 205.1 & $2.5\times 10^{-9}$ & 1 & - & 568.9 & $4.0\times 10^{-6}$ & 2\\
- & 206.8 & $1.5\times 10^{-5}$ & 0 & - & 590.1 & $1.9\times 10^{-6}$ & 2\\
- & 215.8 & $3.9\times 10^{-9}$ & 2 & - & 592.9 & $3.1\times 10^{-7}$ & 2\\
- & 277.6 & $4.3\times 10^{-6}$ & 2 & - & 621.8 & $0.13$ & 1\\
- & 320.3 & $5.3\times 10^{-10}$ & 1 & - & 629.6 & $2.0\times 10^{-5}$ & 1\\
- & 356.8 & $9.5\times 10^{-5}$ & 2 & - & 644.0 & $7.1\times 10^{-9}$ & 1\\
- & 393.3 & $5.1\times 10^{-5}$ & 2 & 658.9(6) & 658.9 & $0.80$ & 0\\
- & 403.2 & $1.2\times 10^{-8}$ & 2 & 663.7(2) & 663.8 & $5.5\times 10^{-3}$ & 2\\
- & 404.5 & $0.024$ & 1 & - & 690.8 & $1.8\times 10^{-6}$ & 2\\
- & 412.2 & $2.2\times 10^{-4}$ & 2 & - & 720.8 & $6.8\times 10^{-11}$ & 2\\
- & 421.9 & $3.6\times 10^{-12}$ & 2 & - & 752.5 & $2.0\times 10^{-6}$ & 1\\
- & 429.4 & $2.4\times 10^{-8}$ & 1 & - & 754.0 & $2.5\times 10^{-14}$ & 2\\
- & 444.0 & $1.8\times 10^{-9}$ & 2 & - & 779.4 & $2.1\times 10^{-5}$ & 1\\
- & 455.8 & $3.9\times 10^{-5}$ & 2 & - & 809.7 & $4.6\times 10^{-12}$ & 1\\
456.1(2) & 456.3 & $5.6\times 10^{-3}$ & 1 & - & 823.2 & $1.9\times 10^{-4}$ & 0\\
- & 462.0 & $0.062$ & 0 & - & 892.8 & $6.3\times 10^{-10}$ & 2\\
- & 466.3 & $2.5\times 10^{-5}$ & 2 & - & 934.3 & $6.6\times 10^{-9}$ & 2\\
- & 473.1 & $6.7\times 10^{-9}$ & 1 & - & 979.9 & $4.9\times 10^{-11}$ & 2\\
- & 479.9 & $2.3\times 10^{-5}$ & 2 & & & &\\
- & 483.5 & $3.9\times 10^{-8}$ & 2 & & & &\\
495.6(5) & 495.3 & $0.15$ & 0 & & & &\\
\end{tabular}
\end{ruledtabular}
\end{table} 

For high-$L$ resonances beyond $L=2$, the resonance widths in K + Rb collisions become orders of magnitude more narrow. For example, MQDT predicts the widest of the $L=4$ resonances to have a width $\lesssim1$ $\mu$G. The time required to perform a FCC calculation of resonances this narrow makes the comparison between the MQDT and FCC calculations challenging. However, predicting the position and width of high-$L$ resonances remains simple within MQDT. Despite only performing a detailed, fully-coupled calculation on a magnetic field grid of spacing $100$ G, our method has found and characterized features $18$ orders of magnitude smaller than this. 

\section{Conclusion}

MQDT has been extended to high partial wave cold collisions. Our specific standardization of reference wave functions has produced a numerically stable calculation of high partial wave MQDT parameters that are smooth in energy and magnetic field. All of these parameters are described by simple power laws at ultracold energies, and accurate expressions for these parameters in the threshold regime have been derived for potentials dominated by $-C_6/R^6$ at long range. As an example, excellent agreement has been shown between MQDT and the FCC calculation of ultracold $^{40}$K + $^{87}$Rb scattering in their lowest hyperfine states. This calculation has also shown good agreement with experimental measurements of Fano-Feshbach resonances, and $\textit{all}$ of the $L=0-2$ Fano-Feshbach resonances in the range of $B=0-1000$ G have been reported.

\section*{ACKNOWLEDGMENTS}
The authors acknowledge financial support from the US Department of Energy.

\begin{subappendices}
\appendix
\section{The Instability of $\phi_i$}
\label{app:tanz}
Here, we consider the calculation of the phase $\phi_i$ that defines the MQDT reference wave functions $\hat{f}$ and $\hat{g}$. $\phi_i$ describes the particular linear combination of $\hat{f}_{(\phi_i=0)}$ and $\hat{g}_{(\phi_i=0)}$ that coincides with a wave function of our choice at zero energy. We choose this wave function according to its asymptotic behavior and define our reference wave functions at $R_\text{x}$. Therefore, we can determine $\phi_i$ by numerically propagating $\hat{f}_{(\phi_i=0)}$ and $\hat{g}_{(\phi_i=0)}$ from their boundary conditions at $R_\text{x}$ to large $R$. However, numerical error causes this propagation to become unstable in the presence of a centrifugal barrier; hence, the numerical calculation of $\phi_i$ can also become unstable. 

For example, we consider the $-1/R^6$ potential. In this potential, every zero-energy wave function is known in terms of the analytically known wave functions (\ref{eq:zero}) at all $R$. These wave functions have well-known asymptotic behaviors and are exact solutions at small $R$. Therefore, their behavior at $R_\text{x}$ determines $\phi_i$ without any numerical propagation. Nevertheless, we can still attempt to determine $\phi_i$ numerically by propagating the wave functions $\hat{f}_{(\phi_i=0)}$ and $\hat{g}_{(\phi_i=0)}$ from their boundary conditions at $R_x$ to large $R$. Then, studying the deviation of our numerically determined value of $\phi_i$ from the analytically known value allows us to characterize the numerical instability of this calculation. Moreover, we identify a robust method for avoiding this instability that is easily generalized to any potential that falls off faster than $1/R^2$ asymptotically.  

We rewrite the zero-energy wave functions $\hat{f}_{(\phi_i=0)}$ and $\hat{g}_{(\phi_i=0)}$ in terms of the analytic wave functions (\ref{eq:zero}) by defining a constant $2 \times 2$ matrix $C$,
\begin{subequations}
\label{eq:cees}
\begin{align}
 \hat{f}_{(\phi_i=0)}&=c_{11} \chi_++c_{12} \chi_-\text{ },\\
 \hat{g}_{(\phi_i=0)}&=c_{21} \chi_++c_{22} \chi_-\text{ }.
\end{align}
\end{subequations}
If we choose to let $\hat{f}$ coincide with $\chi_+$ at zero energy, we can derive an expression for $\tan\phi_i$ in terms of the elements of $C$ by using equation (\ref{eq:tanphi}),
\begin{equation}
 \tan\phi_i=\frac{c_{11}W\left(\chi_+,\chi_+\right)+c_{12}W\left(\chi_+,\chi_-\right)}{c_{21}W\left(\chi_+,\chi_+\right)+c_{22}W\left(\chi_+,\chi_-\right)}=\frac{c_{12}}{c_{22}}\text{ }.
\end{equation}

However, achieving this value of $\tan\phi_i$ numerically is not guaranteed. We can track numerical error by considering the difference between a particular analytic solution and the same solution determined numerically. To this end, we expand the zero-energy wave functions $\chi_+$ and $\chi_-$ in powers of $R$ at large $R$,
\begin{subequations}
\label{eq:analytic}
\begin{align}
\label{eq:analz}
 \chi_+\xrightarrow{R>>1}&\approx R^{L+1}+BR^{L-3}\text{ },\\
 \chi_-\xrightarrow{R>>1}&\approx R^{-L}\text{ },
\end{align}
\end{subequations}
where $B$ is a known constant and the normalization of these wave functions is chosen such that $W(\chi_-,\chi_+)=2L+1$. We then define two numerically determined wave functions $\chi_+'$ and $\chi_-'$ that have the same boundary conditions at $R_\text{x}$ as the analytic solutions $\chi_+$ and $\chi_-$, respectively. Moreover, we demand that $\chi_+$ and $\chi_-$ have the exact same normalization as their analytic counterparts asymptotically.

Numerical error causes the wave functions $\chi_+'$ and $\chi_-'$ to differ from the analytic solutions in two ways. First, only the leading order terms in their asymptotic expansions agree exactly. Hence, the asymptotic expansion of $\chi_+'$ has a coefficient $B'$ in its second highest order term that differs slightly from the coefficient $B$ in equation (\ref{eq:analz}). Thus, we represent our error by the constant $\delta\approx B-B'$. Second, the numerical wave functions become slightly different linear combinations of the analytic solutions, such that $\chi_+'$ gains a contribution from $\chi_-$ that is proportional to the error $\delta$. $\chi_-'$ differs from $\chi_-$ in analogous ways, and we expand both of the numerical wave functions in powers of $R$ at large $R$,
\begin{subequations}
\label{eq:numerical}
\begin{align}
 \chi_+'\xrightarrow{R>>1}&\approx R^{L+1}+B'R^{L-3}+\delta R^{-L}\text{ },\\
 \chi_-'\xrightarrow{R>>1}&\approx R^{-L}+\delta \left( R^{L+1}+BR^{L-3}\right)\text{ }.
\end{align}
\end{subequations}

The presence of numerical error also changes the wave functions $\hat{f}_{(\phi_i=0)}$ and $\hat{g}_{(\phi_i=0)}$. We call these numerically determined wave functions $\hat{f}_{(\phi_i=0)}'$ and $\hat{g}_{(\phi_i=0)}'$, and they lead to the numerically determined phase $\phi_i'$. We define $\hat{f}_{(\phi_i=0)}'$ and $\hat{g}_{(\phi_i=0)}'$ by their boundary conditions at $R_\text{x}$. Therefore, these functions are exactly the functions (\ref{eq:cees}) at $R_\text{x}$, but they take a slightly different form at large $R$. We approximate their large-$R$ behavior as the following,
\begin{align}
   \label{eq:approx}
 \hat{f}_{(\phi_i=0)}'&\xrightarrow{R>>1}c_{11} \chi_+'+c_{12} \chi_-'\text{ },\\
 \hat{g}_{(\phi_i=0)}'&\xrightarrow{R>>1}c_{21} \chi_+'+c_{22} \chi_-'\text{ }.
\end{align}
Using these wave functions in equation (\ref{eq:tanphi}) leads to an equation for $\tan\phi_i'$ that depends on the product $\delta R^{2L-3}$ at large $R$,
\begin{subequations}
\begin{align}
\tan\phi_i'&\xrightarrow{R>>1}\frac{c_{11}W\left(\chi_+,\chi_+'\right)+c_{12}W\left(\chi_+,\chi_-'\right)}{c_{21}W\left(\chi_+,\chi_+'\right)+c_{22}W\left(\chi_+,\chi_-'\right)}\text{ },\\
\label{eq:diverge}
&=\frac{c_{11} 4 \delta R^{2L-3}-c_{12}(2L+1)}{c_{21} 4 \delta R^{2L-3}-c_{22}(2L+1)}\text{ },\\
&\xrightarrow{R\rightarrow\infty}\frac{c_{11}}{c_{21}}\quad\text{ for }L>1\text{ and }\delta\ne0\text{ },
\end{align}
\end{subequations}
For $L>1$, the large-$R$ limit of $\tan\phi_i'$ approaches the wrong value $c_{11}/c_{21}$ if $\delta$ is non-zero. Indeed, performing the actual numerical calculation produces this same value of $\tan\phi_i'$. 

Since the value of $\tan\phi_i$ depends on the constants $c_{12}$ and $c_{22}$ and these terms are dominated by numerical error at large $R$, we deduce that finding the contribution to $\hat{f}_{(\phi_i=0)}$ and $\hat{g}_{(\phi_i=0)}$ from $\chi_-$ at large $R$ is numerically challenging when $L>1$. In fact, letting either $\hat{f}$ or $\hat{g}$ -- at zero energy -- coincide with any wave function with a contribution from $\chi_+$ leads to an equation for $\tan\phi_i$ that depends on the constants $c_{12}$ and $c_{22}$, and the same numerical instability exists. However, if we instead let $\hat{f}$ or $\hat{g}$ coincide with $\chi_-$ at zero energy, using the numerical wave functions $\hat{f}_{(\phi_i=0)}'$ and $\hat{g}_{(\phi_i=0)}'$ at large $R$ leads to an equation for $\tan\phi_i'$ that reduces to the analytic value of $\tan\phi_i$ for all $L$.

For example, if we let $\hat{g}$ coincide with $\chi_-$ at zero energy, we can derive an expression for $\tan\phi_i$ by using the exact values of $\hat{f}_{(\phi_i=0)}$ and $\hat{g}_{(\phi_i=0)}$ in equation (\ref{eq:tanphi2}),
\begin{equation}
 \tan\phi_i=-\frac{c_{21}W\left(\chi_-,\chi_+'\right)+c_{22}W\left(\chi_-,\chi_-'\right)}{c_{11}W\left(\chi_-,\chi_+'\right)+c_{12}W\left(\chi_-,\chi_-'\right)}=-\frac{c_{21}}{c_{11}}\text{ }.
\end{equation}
Using the numerical wave functions $\hat{f}_{(\phi_i=0)}'$ and $\hat{g}_{(\phi_i=0)}'$ at large $R$ leads to the following equations for $\tan\phi_i'$,
\begin{subequations}
\begin{align}
\tan\phi_i'&=-\frac{c_{21}W\left(\chi_-,\chi_+'\right)+c_{22}W\left(\chi_-,\chi_-'\right)}{c_{11}W\left(\chi_-,\chi_+'\right)+c_{12}W\left(\chi_-,\chi_-'\right)}\text{ },\\
&\xrightarrow{R>>1}-\frac{c_{21}+c_{22}\delta }{c_{11}+ c_{12}\delta}\text{ },\\
&\approx-\frac{c_{21}}{c_{11}}\quad\text{ for all }L\text{ and }\delta<<1\text{ }.
\end{align}
\end{subequations}
Here, $\tan\phi_i'$ does not depend on $R$ in the region $R>>1$, and $\tan\phi_i'$ approaches approximately the correct value even if the numerical error is finite. Of course an accurate value of $\tan\phi_i'$ requires the numerical error to be small $\left(\delta<<c_{12} \text{ and } \delta<<c_{22}\right)$, but the divergence seen in equation (\ref{eq:diverge}) does not appear. In this case, $\tan\phi_i$ does not depend on the constants $c_{12}$ and $c_{22}$. Therefore, finding the contribution to $\hat{f}_{(\phi=0)}$ and $\hat{g}_{(\phi=0)}$ from $\chi_-$ at large $R$ is not necessary, and the numerical instability of calculating $\tan\phi_i$ vanishes.


\section{MQDT Threshold Behavior}
\label{app:thresh}
\subsection{Introduction}
The MQDT parameters $A$, $\eta$, $\mathcal{G}$, and $\gamma$ connect the reference wave functions $\hat{f}$ and $\hat{g}$ to well-known solutions in the limit $R\rightarrow\infty$, as described in the equations (\ref{eq:AandG} - \ref{eq:gamma}). By considering a simple long-range potential, we can represent the zero-energy limit of $\hat{f}$ and $\hat{g}$ in terms of analytically known zero-energy solutions. However, these solutions are inadequate to describe the large-$R$ behavior of $\hat{f}$ and $\hat{g}$ at non-zero energies, so we find a correction to the zero-energy wave functions via perturbation theory. With an accurate representation of $\hat{f}$ and $\hat{g}$ at small energies in hand, we derive simple expressions for the MQDT parameter threshold behavior by matching these wave functions to either the energy-dependent wave functions $f$ and $g$ or the function $e^{-\kappa R}$.
\subsection{Zero-Energy Solutions}
For the simple long-range potential $-C_6/R^6$, we can solve the Schr\"{o}dinger equation,
\begin{equation}
\label{eq:appSE}
 -\frac{\mathrm{d}^2\psi}{\mathrm{d}R^2}+\frac{L(L+1)\psi}{R^2}-\frac{\psi}{R^6}=E\psi\text{ },
\end{equation}
analytically at $E=0$. For all of Appendix \ref{app:thresh}, $R$ is in units of the natural length scale $\beta=(2\mu C_6/\hbar^2)^{1/4}$ of the potential $-C_6/R^6$, and $E$ is in units of the natural energy scale $E_{\beta}=\hbar^2/2\mu\beta^2$, where $\mu$ is the reduced mass. We describe particular solutions to equation (\ref{eq:appSE}) at zero energy in terms of two linearly independent solutions $\chi_+$ and $\chi_-$ defined by their asymptotic behavior,
\begin{subequations}
\label{eq:chis}
\begin{align}
 \chi_+&=\sqrt{R}J_{-\frac{1}{4}(2L+1)}(1/2R^2)\xrightarrow{R\rightarrow\infty}\frac{2^{L+1/2}R^{L+1}}{\Gamma(3/4-L/2)} \text{ },\\
 \chi_-&=\sqrt{R}J_{\frac{1}{4}(2L+1)}(1/2R^2)\xrightarrow{R\rightarrow\infty}\frac{2^{-(L+1/2)}R^{-L}}{\Gamma(L/2+5/4)}\text{ },
\end{align}
\end{subequations}
where $J$ is the Bessel function of the first kind.

For all energies, we define two linearly independent reference wave functions $\hat{f}$ and $\hat{g}$ with the following boundary conditions at $R_\text{x}<<1$,
\begin{subequations}
\label{eq:refs}
\begin{align}
\label{eq:fhat}
\hat{f}(R)&=\frac{1}{\sqrt{k(R)}}\sin(\int_{R_\text{x}}^{R} k(R') \mathrm{d}R' + \phi) \quad\text{at }R=R_\text{x}\text{ },\\
\label{eq:ghat}
\hat{g}(R)&=\frac{-1}{\sqrt{k(R)}}\cos(\int_{R_\text{x}}^{R} k(R') \mathrm{d}R' + \phi) \quad\text{at }R=R_\text{x}\text{ }.
\end{align}
\end{subequations}
Here, $\phi$ is a phase that is constant in $R$ and energy, and $k=\sqrt{E+1/R^6}$. The set of equations (\ref{eq:refs}) and their full radial derivatives define $\hat{f}$ and $\hat{g}$.

We then demand that $\hat{g}$ coincides (up to a normalization) with the solution $\chi_-$ at zero energy. Hence, we rewrite $\hat{f}$ and $\hat{g}$ in terms of $\chi_+$ and $\chi_-$ at zero energy by defining two constants of normalization $N_1$ and $N_2$ and a constant phase $\alpha$,
\begin{subequations}
\label{eq:constants}
\begin{align}
\label{eq:fzero}
 \hat{f}(E=0)&=N_2\left(\chi_++\tan\alpha \chi_-\right)\text{ }.\\
 \label{eq:gzero}
 \hat{g}(E=0)&=N_1\chi_-\text{ }.
\end{align}
\end{subequations}
By considering the small-$R$ limit of our zero-energy solutions and reference wave functions,
\begin{subequations}
\begin{align}
 \chi_+&\xrightarrow{R<<1}\frac{2}{\sqrt{\pi}}R^{3/2}\sin\left(-\frac{1}{2R^2}-\frac{L \pi}{4}+\frac{5\pi}{8}\right)\text{ },\\
 \chi_-&\xrightarrow{R<<1}-\frac{2}{\sqrt{\pi}}R^{3/2}\cos\left(-\frac{1}{2R^2}+\frac{L \pi}{4}-\frac{5\pi}{8}\right)\text{ },
\end{align}\begin{align}
 \hat{f}(E=0)&\xrightarrow{R<<1}R^{3/2}\sin\left(-\frac{1}{2R^2}+\frac{1}{2R_\text{x}^2}+\phi\right)\text{ },\\
 \hat{g}(E=0)&\xrightarrow{R<<1}-R^{3/2}\cos\left(-\frac{1}{2R^2}+\frac{1}{2R_\text{x}^2}+\phi\right)\text{ },
\end{align} 
\end{subequations}
we use the sets of equations (\ref{eq:refs}) and (\ref{eq:constants}) to determine the four unknown constants,
\begin{subequations}
 \begin{align}
  N_1&=\frac{\sqrt{\pi}}{2}\text{ },\\
 \phi&=-\frac{1}{2R_\text{x}^2}+\frac{L\pi}{4}-\frac{5\pi}{8}\text{ },\\
 \tan\alpha&=(-1)^{L+1}\sin\left(\frac{2L+1}{4}\pi\right)\text{ },\\
  N_2&=-\frac{\sqrt{\pi}}{2\sin\left(\frac{2L+1}{4}\pi\right)}\text{ }.
 \end{align}
\end{subequations}

\subsection{Perturbation Theory}
At zero energy, we know the wave functions $\hat{f}$ and $\hat{g}$ exactly; however, it is not immediately obvious whether or not the zero-energy wave functions are good approximations at large $R>>1$ in the limit of $E\rightarrow0$. At small energies $E<<1$, both $\hat{f}$ and $\hat{g}$ grow with $R$ before reaching their asymptotic limits, but only $\hat{f}$ grows at exactly zero energy. Since $\hat{f}$ has a contribution from $\chi_+$ and $\chi_-$ at zero energy, matching to finite-energy wave functions is straightforward. However, $\hat{g}$ is purely $\chi_-$ at zero energy, and the contribution to $\hat{g}$ from $\chi_+$ at small energies is unknown. In order to match our zero-energy wave functions onto growing, finite-energy wave functions at large $R$, we must find the contribution to $\hat{g}$ from $\chi_+$. This is accomplished by performing a perturbation in $E$. 

Because we plan to match wave functions at a finite $R$, we choose a Green's function which preserves the boundary conditions of $\hat{g}$ at $R_\text{x}$ \cite{Taylor2006},
\begin{equation}
 G(R,R')=
\begin{cases}
 0&\text{ if }R<R'\text{ },\\
 \frac{\chi_+(R)\chi_-(R')-\chi_-(R)\chi_+(R')}{(N_1N_2)^{-1}}&\text{ if }R>R'\text{ }.
\end{cases}
\end{equation}
Hence, there is an integral equation for $\hat{g}$ at small energies,
\begin{widetext}
\begin{subequations}
\begin{align}
\hat{g}(R,E<<1)&=\hat{g}(R,E=0)+\int_0^RG(R,R')\hat{g}(R',E=0)\mathrm{d}R'\text{ }, \\
\label{eq:ghatanal}
&=N_1\chi_-(R)+E N_1^2N_2\bigg(\chi_+(R)\int_0^R \chi_-^2(R')\mathrm{d}R'-\chi_-(R)\int_0^R \chi_+(R')\chi_-(R')\mathrm{d}R'\bigg)\text{ },
\end{align}
\end{subequations}
where we have used equation (\ref{eq:gzero}) to replace $\hat{g}$ at zero energy. We solve these integrals analytically and then expand them in powers of $R$ at large $R$,
\begin{subequations}
 \begin{align}
  \int_0^R \chi_-^2(R')\mathrm{d}R'&\xrightarrow{R>>1}\frac{4}{(2L+3)(2L-1)\pi}-\frac{2^{-(2L+1)}R^{1-2L}}{(2L-1)\Gamma(5/4+L/2)^2}\text{ },\\
  \int_0^R \chi_-(R')\chi_+(R')\mathrm{d}R'&\xrightarrow{R>>1}\frac{4 \cos\left(\frac{2L+1}{4}\pi\right)}{(2L+3)(2L-1)\pi}+\frac{2\sin\left(\frac{2L+1}{4}\pi\right)R^2}{(2L+1)\pi}\text{ }.
 \end{align}
\end{subequations}
By using this correction to the zero-energy $\hat{g}$ and approximating $\hat{f}$ with its zero-energy limit, we have complete descriptions of the reference wave functions at large $R$ in the limit of zero energy,
\begin{subequations}
\label{eq:refexp}
\begin{align}
\label{eq:fhatexp}
  \hat{f}&\xrightarrow[E\rightarrow0]{R>>1} N_2\left(\frac{2^{L+1/2}R^{L+1}}{\Gamma(3/4-L/2)}+\tan\alpha\frac{2^{-(L+1/2)}R^{-L}}{\Gamma(5/4+L/2)}\right)\text{ },\\
\label{eq:ghatexp}
  \hat{g}&\xrightarrow[E\rightarrow0]{R>>1} N_1\frac{2^{-(L+1/2)}R^{-L}}{\Gamma(5/4+L/2)}\nonumber\\
&+E N_1^2N_2\left(\frac{2^{L+5/2}R^{L+1}}{\Gamma(3/4-L/2)(2L+3)(2L-1)\pi}-\frac{2^{-L+1/2}\sin\left(\frac{2L+1}{4}\pi\right)R^{-L+2}}{\Gamma(5/4+L/2)(2L-1)\pi}-\frac{2^{-L+3/2}\cos(\frac{2L+1}{4}\pi)R^{-L}}{\Gamma(5/4+L/2)(2L+3)(2L-1)\pi}\right)\text{ }.
 \end{align}
\end{subequations}
\end{widetext}
\subsection{Matching Wave Functions}
The MQDT parameters connect $\hat{f}$ and $\hat{g}$ with $f$, $g$, and $e^{-\kappa R}$. Since we have analytic expressions for all of these wave functions at large $R$ and small energies, we use equations (\ref{eq:AandG} - \ref{eq:gamma}) to solve for the threshold behavior of the MQDT parameters. Moreover, by expanding these wave functions in powers of $R$ and comparing like terms, we derive simple formulas. We can use the asymptotic expansions of $\hat{f}$ and $\hat{g}$ at small energies in equations (\ref{eq:fhatexp}) and (\ref{eq:ghatexp}), but we still need to find similar expansions of $f$ and $g$ in this same parameter regime: $R>>1$ and $E\rightarrow0$. 

At large $R$, we rewrite $f$ and $g$ in terms of spherical Bessel functions using (\ref{eq:bess}) and (\ref{eq:fandg}),
\begin{subequations}
\begin{align}
 f&\xrightarrow{R\rightarrow\infty}\frac{kR}{\sqrt{k}}\left(j(kR)\cos\eta-n(kR)\sin\eta\right)\text{ },\\
 g&\xrightarrow{R\rightarrow\infty}\frac{kR}{\sqrt{k}}\left(n(kR)\cos\eta+j(kR)\sin\eta\right)\text{ }.
\end{align}
\end{subequations}
Then, the small-argument expansions of the spherical Bessel functions unveil the behavior of $f$ and $g$ at large $R$ and very small energies such that $kR<<1$,
\begin{subequations}
\label{eq:fandgexp}
 \begin{align}
\label{eq:fexp}
f&\xrightarrow[k\rightarrow0]{R>>1}\frac{1}{\sqrt{k}}\frac{(kR)^{L+1}}{(2L+1)!!}\cos\eta+\frac{1}{\sqrt{k}}\frac{(2L-1)!!}{(kR)^L}\sin\eta\text{ },\\ 
\label{eq:gexp}
g&\xrightarrow[k\rightarrow0]{R>>1}-\frac{1}{\sqrt{k}}\frac{(2L-1)!!}{(kR)^L}\cos\eta\left(1+\frac{(kR)^2}{4L-2}\right)\nonumber\\
&+\frac{1}{\sqrt{k}}\frac{(kR)^{L+1}}{(2L+1)!!}\sin\eta\text{ },
 \end{align}
\end{subequations}
where $k=\sqrt{E}$. We see that $f$ has a term proportional to $R^{L+1}$ and a term proportional to $R^{-L}$. Hence, we can compare this function with $\hat{f}$ term by term. The equation for $f$ in (\ref{eq:AandG}),
\begin{equation}
 A^{-1/2}f=\hat{f}\text{ },
\end{equation}
yields two equations for the MQDT parameters $A$ and $\eta$, 
\begin{subequations}
\label{eq:Aandeta}
 \begin{align}
  \label{eq:A}
  A^{1/2}&=\frac{\Gamma(3/4-L/2)k^{L+1/2}\cos\eta}{N_2 2^{L+1/2} (2L+1)!!}\text{ },\\
  \label{eq:eta}
  \sin\eta&=\frac{N_2\tan\alpha 2^{-(L+1/2)}A^{1/2}k^{L+1/2}}{\Gamma(5/4+L/2)(2L-1)!!}\text{ },
 \end{align}
\end{subequations}
where matching powers of $R^{L+1}$ leads to equation (\ref{eq:A}) and matching powers of $R^{-L}$ leads to equation (\ref{eq:eta}).

From (\ref{eq:Aandeta}) we find that $\tan\eta\propto k^{2L+1}$. Thus, for small $k$, we use the small-angle approximation, $\sin\eta\approx\eta$ and $\cos\eta\approx 1$, and define the generalized scattering length $\bar{a}_L$,
\begin{equation}
 \bar{a}_L=\left(\frac{\pi 2^{-(2L+3/2)}}{\Gamma(L/2+5/4)\Gamma(L+1/2)}\right)^{2/(2L+1)}\text{ }.
\end{equation}
We then rewrite our expressions for $A$ and $\eta$ in terms of $\bar{a}_L$, 
\begin{subequations}
\begin{align}
\label{eq:Athresh}
 A^{1/2}&=-(\bar{a}_Lk)^{L+1/2}\text{ },\\
\eta&=(-1)^{L+1}(\bar{a}_Lk)^{2L+1}\text{ }.
\end{align}
\end{subequations}
Here, we have used the relations $(2n-1)!!=2^{n}\Gamma(1/2+n)/\sqrt{\pi}$, where $n$ is an integer, and $\sin\left(\pi z\right)=\pi/\Gamma(1-z)\Gamma(z)$ with $z=(2L+1)/4$. For $L>1$ and small $k$, we know the phase shift is dominated by a long-range phase shift proportional to $k^4$ \cite{Landau1977}, but the derivation above only yields the short-range contribution because we are matching wave functions under the centrifugal barrier. Since we know the long-range contribution analytically, we simply correct our expression for $\eta$ by adding these contributions together,
\begin{equation}
 \eta=(-1)^{L+1}(\bar{a}_Lk)^{2L+1}+\frac{3 \pi \Gamma(L-3/2)}{32\Gamma(L+7/2)}k^4\text{ }.
\end{equation}

The length scale $\bar{a}_L$ helps to greatly reduce the amount of unnecessary constants in the derivation of the remaining MQDT parameters $\mathcal{G}$ and $\gamma$. Hence, we rewrite our wave functions in terms of $\bar{a}_L$. At large $R$, the zero-energy solutions become
\begin{subequations}
\begin{align}
 \chi_+&\xrightarrow{R>>1}-\frac{1}{N_2}\sqrt{\bar{a}_L}\frac{\left(R/\bar{a}_L\right)^{L+1}}{(2L+1)!!}\text{ },\\
 \chi_-&\xrightarrow{R>>1}\frac{1}{N_1}\sqrt{\bar{a}_L}\frac{(2L-1)!!}{\left(R/\bar{a}_L\right)^{L}}\text{ }, 
\end{align}
\end{subequations}
and we have simple expressions for the zero-energy limits of $\hat{f}$ and $\hat{g}$ at large $R$,
\begin{widetext}
\begin{subequations}
\label{eq:refexp2}
\begin{align}
 \hat{f}&\xrightarrow[E\rightarrow0]{R>>1} -\sqrt{\bar{a}_L}\frac{\left(R/\bar{a}_L\right)^{L+1}}{(2L+1)!!}+(-1)^L\sqrt{\bar{a}_L}\frac{(2L-1)!!}{\left(R/\bar{a}_L\right)^{L}}\text{ },\\
\label{eq:ghatexp2}
 \hat{g}&\xrightarrow[E\rightarrow0]{R>>1}\sqrt{\bar{a}_L}\frac{(2L-1)!!}{\left(R/\bar{a}_L\right)^{L}}\left(1+\frac{E R^2}{4L-2}+\frac{(-1)^LE}{(2L+3)(2L-1)}\right)-\frac{E}{(2L+3)(2L-1)}\sqrt{\bar{a}_L}\frac{\left(R/\bar{a}_L\right)^{L+1}}{(2L+1)!!}\text{ }.
\end{align}
\end{subequations}

We derive the threshold behavior of the MQDT parameter $\mathcal{G}$ by using the equation for $g$ in (\ref{eq:AandG}),
\begin{equation}
\label{eq:getG}
 A^{1/2}g=\hat{g}+\mathcal{G}\hat{f}\text{ }.
\end{equation}
In this equation, we substitute $\hat{g}$ and $g$ with their expansions in (\ref{eq:ghatexp2}) and (\ref{eq:gexp}), respectively, and replace $A$ by its threshold value in equation (\ref{eq:Athresh}). Thus, in the limit of large $R$ and very small, positive energy such that $kR<<1$, we evaluate the LHS and RHS of equation (\ref{eq:getG}) separately,
\begin{subequations}
\begin{align}
 A^{1/2}g&\xrightarrow[k\rightarrow0]{R>>1}\sqrt{\bar{a}_L}\frac{(2L-1)!!}{(R/\bar{a}_L)^L}\left(1+\frac{(kR)^2}{4L-2} \right) +(-1)^L(\bar{a}_Lk)^{4L+2}\sqrt{\bar{a}_L}\frac{(R/\bar{a}_L)^{L+1}}{(2L+1)!!}\text{ },\\
\hat{g}+\mathcal{G}\hat{f}&\xrightarrow[k\rightarrow0]{R>>1}\sqrt{\bar{a}_L}\frac{(2L-1)!!}{\left(R/\bar{a}_L\right)^{L}}\left(1+\frac{(kR)^2}{4L-2}+\frac{(-1)^Lk^2}{(2L+3)(2L-1)}+(-1)^L\mathcal{G}\right)-\sqrt{\bar{a}_L}\frac{\left(R/\bar{a}_L\right)^{L+1}}{(2L+1)!!}\left(\frac{k^2}{(2L+3)(2L-1)}+\mathcal{G}\right)\text{ },
\end{align}
\end{subequations}
where $E=k^2$. Hence, the first two terms on the LHS cancel exactly with the first two terms on the RHS, leading to the following equation,
\begin{equation}
(-1)^L(\bar{a}_Lk)^{4L+2}\frac{(R/\bar{a}_L)^{L+1}}{(2L+1)!!}
=-\frac{\left(R/\bar{a}_L\right)^{L+1}}{(2L+1)!!}\left(\frac{k^2}{(2L+3)(2L-1)}+\mathcal{G}\right)+(-1)^L\frac{(2L-1)!!}{\left(R/\bar{a}_L\right)^{L}}\left(\frac{k^2}{(2L+3)(2L-1)}+\mathcal{G}\right)\text{ }.
\end{equation}
\end{widetext}
In the limit of large $R>>1$, the $R^{L+1}$ terms on the RHS dominate the $R^{-L}$ terms for all $L$ and all $k$, independent of $\mathcal{G}$; therefore, neglecting the terms of order $R^{-L}$ in this equation gives the threshold behavior of $\mathcal{G}$,
\begin{equation}
 \mathcal{G}=(-1)^{L+1}(\bar{a}_Lk)^{4L+2}-\frac{k^2}{(2L+3)(2L-1)}\text{ }.
\end{equation}


We derive the threshold behavior of $\gamma$ in a way similar to the derivation of $\mathcal{G}$. Here, instead of matching to $g$, we need to match the small-energy limit of $\hat{f}$ and $\hat{g}$ to the function $e^{-\kappa R}$ at large $R$. We again try to match wave functions at large $R$ and very small energies such that $\kappa R<<1$. Using equation (\ref{eq:gamma}), we define a constant of proportionality $D$,
\begin{equation}
\label{eq:tangamma}
  \tan\gamma_i\hat{f}_i+\hat{g}_i\xrightarrow{R>>1}D e^{-\kappa_iR}\text{ }.
\end{equation}
Then, using (\ref{eq:refexp2}) for $\hat{f}$ and $\hat{g}$ and using the $\kappa R<<1$ expansion of $e^{-\kappa R}$, we arrive at the following equation,
\begin{widetext}
\begin{align}
D\sum_{n=0}^\infty\frac{(-\kappa R)^n}{n!}&= -\sqrt{\bar{a}_L}\frac{\left(R/\bar{a}_L\right)^{L+1}}{(2L+1)!!}\left(-\frac{\kappa^2}{(2L+3)(2L-1)}+\tan\gamma\right)\nonumber\\
\label{eq:findgamma}
&+\sqrt{\bar{a}_L}\frac{(2L-1)!!}{\left(R/\bar{a}_L\right)^{L}}\left(1-\frac{(\kappa R)^2}{4L-2}-\frac{(-1)^L\kappa^2}{(2L+3)(2L-1)}+(-1)^L\tan\gamma\right)\text{ },
\end{align}
\end{widetext}
where $E=-\kappa^2$.

For $L=0$, we take the expansion of $e^{-\kappa R}$ out to first order in $\kappa$ ($n=1$) and neglect terms of order $\kappa^2$,
\begin{equation}
 D(1 -\kappa R)=\sqrt{\bar{a}_0}\left(1+\tan\gamma-\tan\gamma R/\bar{a}_0\right)\text{ }.
\end{equation}
Matching constant terms and terms of order $R$ leads to the following two equations with two unknowns,
\begin{subequations}
 \begin{align}
  D&=\sqrt{\bar{a}_0}\left(1+\tan\gamma\right)\text{ },\\
  -D\kappa &=-\tan\gamma/\sqrt{\bar{a}_0}\text{ }.
 \end{align}
Hence,
\begin{align}
 D&=\frac{\tan\gamma}{\sqrt{\bar{a}_0}\kappa}\text{ },\\
 \tan\gamma&=\frac{1}{1-\bar{a}_0\kappa}\text{ },
\end{align}
\end{subequations}
and we have a simple formula for $\tan\gamma$ in the threshold limit $\kappa<<1$ when $L=0$,
\begin{equation}
 \tan\gamma=\bar{a}_0\kappa\quad\text{ for }L=0\text{ }.
\end{equation}

For $L>0$, we immediately see that matching powers of $R$ in equation (\ref{eq:findgamma}) is problematic due to the terms of order $R^{-L}$ on the RHS. Therefore, instead of matching in the limit $\kappa R<<1$, we simply let the wave functions take their asymptotic forms as $R\rightarrow\infty$, where $\kappa R>>1$ even though $\kappa<<1$. That is, $e^{-\kappa R}\rightarrow0$, and $\hat{f}$ and $\hat{g}$ are still well approximated by their $E\rightarrow0$ limits in (\ref{eq:refexp2}). As $R$ becomes very large, these wave functions are dominated by their contributions from $R^{L+1}$. Even the term in $\hat{g}$ proportional to $R^{-L+2}$ is dominated by $R^{L+1}$ for $L>0$, and equation (\ref{eq:tangamma}) takes a simple form,
\begin{equation}
 0= -\sqrt{\bar{a}_L}\frac{\left(R/\bar{a}_L\right)^{L+1}}{(2L+1)!!}\left(-\frac{\kappa^2}{(2L+3)(2L-1)}+\tan\gamma\right)\text{ }.
\end{equation}
Hence, we have the following simple formulas for the threshold behavior of $\gamma$,
\begin{equation}
 \gamma=
\begin{cases}
 \bar{a}_0\kappa&\text{for }L=0\text{ },\\
 \frac{\kappa^2}{(2L+3)(2L-1)}&\text{for }L>0\text{ }.
\end{cases}
\end{equation}
\end{subappendices}

\bibliographystyle{myapsrev.bst}
\bibliography{paperdb.bib}
\end{document}